\documentclass[prb,twocolumn,showpacs,floats,eqsecnum,amsmath,amssymb,nofootinbib]{revtex4}
\usepackage[dvips]{graphicx}

\begin{document}



\title{Real space renormalization group approach to the two-dimensional 
antiferromagnetic 
Heisenberg model (I)
- The singlet triplet gap.}

\author{A. Fledderjohann$^{1}$, A. Kl\"umper$^{1}$ and K.-H. M\"utter$^{1}$}

\affiliation{$^1$Physics Department, University of Wuppertal, 42097 Wuppertal,
Germany}


\begin{abstract}
\leftskip 2cm
\rightskip 2cm
The low energy behaviour of the two-dimensional antiferromagnetic Heisenberg model is
studied in the sector with total spins $S=0,1,2$ by means of a renormalization
group procedure, which generates a recursion formula for the interaction
matrix $\Delta_S^{(n+1)}$ of 4 neighbouring ``$n$ clusters'' of size
$2^n\times 2^n$, $n=1,2,3,\ldots$ from the corresponding quantities 
$\Delta_S^{(n)}$. Conservation of total spin $S$ is implemented explicitly
and plays an important role. It is shown, how the ground state energies
$E_S^{(n+1)}$, $S=0,1,2$ approach each other for increasing $n$, i.e. system
size.
The most relevant couplings in the interaction matrices are generated by the
transitions $\langle S',m';n+1|S_q^*|S,m;n+1\rangle$ between the ground
states $|S,m;n+1\rangle$ ($m=-S,\ldots,S$) on an $(n+1)$-cluster of size
$2^{n+1}\times 2^{n+1}$, mediated by the staggered
spin operator $S_q^*$.
\end{abstract}


\maketitle



\section{Introduction\label{sec1}}
\setcounter{equation}{0}

The discovery of high $T_c$ superconductors 20 years ago \cite{bednorz86}
led to an intensive search for new mechanisms, which could explain the
observed superconductivity in $CuO_2$ planes. It became clear very soon,
that two ingredients are needed, holes in low concentration which move
in a $2D$ antiferromagnetic background. 
The Hubbard model\cite{hubbard63} and the $t-J$ model\cite{anderson87}
have been discussed in this context. If the hopping of holes is frozen in,
these models reduce to the Heisenberg model with Hamiltonian
\begin{eqnarray}
H & = & \sum_{\langle x,y\rangle}{\bf S}(x){\bf S}(y)
\label{H}
\end{eqnarray}
where $S_i(x)$, $i=1,2,3$ denote spin $1/2$ operators at site $x$.
In the absence of an external field (\ref{H}) conserves the total spin
\begin{eqnarray}
{\bf S} & = & \sum_x {\bf S}(x)
\end{eqnarray}
and the ground state is known to be a singlet state ($S=0$) with
momentum ${\bf p}=(0,0)$.
The first excited state $|1,q\rangle$ is a triplet ($S=1,\,\,q=0,\pm 1$)
with momentum ${\bf p}=(\pi,\pi)$. The transition between these two
states is mediated by the staggered spin operator
\begin{eqnarray}
S_q^* & = & \sum_x(-1)^xS_q(x)\hspace{0.8cm}q=0,\pm 1\\
S_0(x) & = & \frac{1}{2}\sigma_3(x)\,,\hspace{0.6cm}
S_{\pm}(x)=\frac{1}{2\sqrt{2}}\Big(\sigma_1(x)\pm i\sigma_2(x)\Big)\nonumber\\
 & & \label{ST_operators}
\end{eqnarray}
and the matrix element
\begin{eqnarray}
m^* & = & \frac{1}{N}\langle 1q|S_q^*|0\rangle
\label{m_staggered}
\end{eqnarray}
can be considered as an order parameter.

The ground state properties of the $2D$ Heisenberg model (\ref{H}) have
been investigated with various methods. The variational $RVB$-state
\cite{baskaran87}
(``Resonating Valence Bond'') starts from singlet states on pairs of
sites, which cover the whole $2D$ lattice. By construction, these
states lead to eigenstates of the total spin ${\bf S}$ with $S=0$.
However, the manifold of these states which can be constructed
is nonorthogonal and overcomplete.

Numerical methods - e.g. the Lanczos algorithm - are limited to small
clusters $N\leq 6\times 6=36$ due to storage problems. The computation
on the largest cluster $6\times 6$ has been performed by Schulz and
Ziman\cite{schulz92} 15 years ago.

In spite of the great improvements achieved during this time, it 
is not possible so far to repeat the calculation of Schulz and Ziman
for the next interesting cluster $8\times 8$.
This is only possible with other techniques, 
as the quantum Monte Carlo [cf. e.g. (\onlinecite{manousakis91})].


Our approach to the low energy properties of the two-dimensional antiferromagnetic
Heisenberg model starts from a decomposition of the lattice into
plaquettes as depicted in Fig. \ref{fig_1}.
Quantum numbers, energies and couplings for plaquette states are 
discussed in Section \ref{sec2}. In particular we find, that the
lowest energy plaquette states with total spin $S=0$ (singlet),
$S=1$ (triplet), $S=2$ (quintuplet) appear to be most important
for the construction of the ground states on larger clusters.
This is demonstrated in Section \ref{sec3} where we compose 4
plaquettes to a $4\times 4$ cluster with open boundary conditions
(cf. Fig. \ref{fig_2}). We deduce interaction matrices from the couplings between
neighbouring plaquettes for the $4\times 4$
system in the sectors with total spin $S=0,1,2$.
The diagonalization of these interaction matrices yield predictions
for energies and transition matrix elements, which can be compared
with Lanczos results on a $4\times 4$ system. We find agreement
within $1-6\%$, depending on the quantity under consideration.

In a next step, described in Section \ref{sec4}, we generalize this
approach to larger ``$n$-clusters'' of size $2^n\times 2^n$,
$n=2,3,4,\ldots$. E.g. the $n=3$ cluster of size $8\times 8$ is
composed from four $n=2$ clusters ($4\times 4$). Each of these 
clusters can be occupied with a cluster ground state with total
spin $S$ ($S=0,1,2$). The interaction matrix for the
$n+1$-clusters have the same structure as in the step before -
i.e. for $n$-clusters. The $n$-dependence can be absorbed in a
renormalization of couplings and gaps, which is discussed in
Sections \ref{sec5} and \ref{sec6}.
The numerical evaluation of the renormalization group equations
is discussed in Section \ref{sec7}. Section \ref{sec8} is devoted
to the treatment of the staggered magnetization in our approach.



Let us finally mention, that the idea to describe the ground state
properties of the antiferromagnetic Heisenberg model in the framework
of a renormalization group approach is not new. One of the
first attempts in this direction has been presented already in
1992 by Lin and Campbell.\cite{lin92,lin94} They started from $L\times L$
clusters with $L$ odd ($L=3,5$) and computed the ground state (which
has total spin 1/2) and its interaction with sites on a ring. In 
this way, they were able to make a prediction for the staggered
magnetization on a $7\times 7$ lattice.

There have been many investigations of ordered antiferromagnets, which
start from the observation that the dominant fluctuations are controlled
by the quantum nonlinear $\sigma$-model with imaginary time.
\cite{chakravarty88,fisher89,affleck86,affleck87}

The various renormalization group approaches differ in the clusters used
and in the truncation of the Hilbert space, which is needed to make
the evaluation feasible. This is discussed in Section \ref{sec9}.

\section{Plaquette states: Quantum numbers, energies and couplings\label{sec2}}
\setcounter{equation}{0}

\begin{figure}[ht!]
\centerline{\hspace{0.0cm}\includegraphics[width=3.5cm,angle=-90]{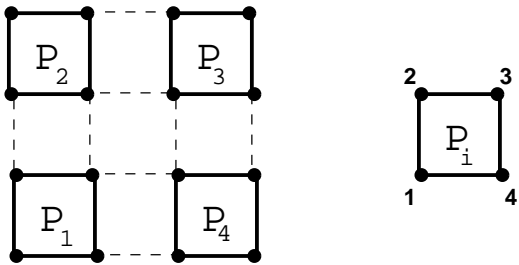}
\hspace{0.0cm}}
\caption{4-plaquette system with $2^n\times 2^n=4\times 4$ sites -- here $n=2$;
the single plaquette on the right shows the enumeration of plaquette sites.}
\label{fig_1}
\end{figure}

Our approach to the $2D$ antiferromagnetic Heisenberg model starts with a 
decomposition of the $2D$ lattice into plaquettes as depicted in Fig. 
\ref{fig_1}.
Each plaquette can carry 16 eigenstates. Their quantum numbers and energies
are listed in Table \ref{table1}.

\begin{table}[ht!]
\begin{tabular}{c|c|c|c|c|c|c}
 & $|0\rangle$ & $|A_q\rangle$ & $|B_q\rangle$ & $|C_q\rangle$ & $|\hat 0\rangle$ & 
$|Q_r\rangle$ \\ \hline
 $S(P)$ & 0 & 1 & 1 & 1 & 0 & 2\\ \hline
$S_z(P)$ & 0 & $q=\pm 1,0$ & $q=\pm 1,0$ & $q=\pm 1,0$ & 0 & $r=\pm 2,\pm 1,0$\\ \hline
$E(P)$ & -2 & -1 & 0 & 0 & 0 & 1
\end{tabular}
\caption{Quantum numbers of the 16 plaquette states.}
\label{table1}
\end{table}

The spin quantum numbers $S(P)$ in the first row result from a decomposition of
the 4 spins into irreducible representations of the $SU(2)$. In this way we get
two singlets $|0\rangle$, $|\hat 0\rangle$ with energies $E_0=-2$, $E_{\hat 0}=0$,
three triplets $|Aq\rangle$, $Bq\rangle$, $|Cq\rangle$ with energies $E_A=-1$,
$E_B=E_C=0$ and one quintuplet $|Q_r\rangle$ with energy $E_2=1$.

The ground state $|0\rangle$ is given by the following spin configuration on the
plaquette

\begin{eqnarray}
|0\rangle & = &
\frac{1}{\sqrt{12}}\left\{\left(\begin{array}{cc} + & +\\- & -
\end{array}\right)+
\left(\begin{array}{cc} - & -\\+ & +\end{array}\right)+
\left(\begin{array}{cc} + & -\\+ & -\end{array}\right)+\right.\nonumber\\
  & & \left. \hspace{0.9cm}
\left(\begin{array}{cc} - & +\\- & +\end{array}\right)-2
\left(\begin{array}{cc} + & -\\- & +\end{array}\right)-2
\left(\begin{array}{cc} - & +\\+ & -\end{array}\right)
\right\}\,.\nonumber\\
\label{GS_0}
\end{eqnarray}
The three triplet states are obtained 
\begin{eqnarray}
|i,q\rangle & = & S_q^{(i)}|0\rangle\frac{1}{\langle 0|S_{-q}^{(i)}S_q^{(i)}
|0\rangle^{1/2}}\quad i=A,B,C\quad\quad
\end{eqnarray}
by application of the plaquette spin operators
\begin{eqnarray}
S_q^{(i)} & = & \sum_{x\in P}\chi^{(i)}(x)S_q(x)\quad q=0,\pm 1\quad
i=A,B,C\nonumber\\
 & & \label{staggered}
\end{eqnarray}
on the ground state (\ref{GS_0}). The signs $\chi^{(i)}(x)$
\begin{eqnarray}
\chi^{(A)}(x)=(+-+-), & \quad & \chi^{(B)}(x)=(++--),\quad\nonumber\\
\chi^{(C)}(x)=(+--+)\, & &
\end{eqnarray}
define the magnetic order of triplet states on the plaquette:

$|A_q\rangle$ is antiferromagnetic in the sense, that it changes sign running around the plaquette

$|B_q\rangle$ and $|C_q\rangle$ are collinear antiferromagnetic in horizontal and
vertical direction, respectively.

The tensor states $|Q_r\rangle$, $r=-2,-1,0,1,2$ are totally symmetric with
respect to the four sites of the plaquette.

The simplest variational ansatz on the $2D$ lattice would start from a product
state   where all plaquettes  are occupied with singlets. Such an ansatz would
lead to an energy per plaquette $E_0(P)=-2$ which is just $75\%$ of the ``exact''
value
\begin{eqnarray}
\hat E_0 & = & 4\cdot(-0.668)=-2.674
\end{eqnarray}
as it follows for the thermodynamical limit from a finite-size scaling analysis
[Huse\cite{huse88} (1988)]. Recent quantum Monte Carlo calculations [Sandvik\cite{
sandvik97} (1997), Loew\cite{loew07} (2007)] 
improve the ground state energy to $e_0=0.669437(5)$.

\begin{figure}[ht!]
\centerline{\hspace{0.0cm}\includegraphics[width=7.5cm,angle=0]{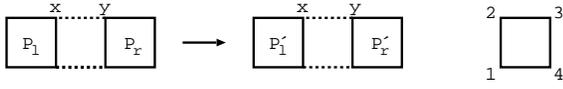}
\hspace{0.0cm}}
\caption{Interaction of neighbouring plaquettes.
}
\label{fig_2}
\end{figure}

Therefore the interaction between neighbouring plaquettes $P_l$ $P_r$ - as
depicted in Fig. \ref{fig_2} - has to account for 25$\%$ of the ground state
energy.
If the two neighbouring plaquettes carry spins $|S_l,m_l\rangle$, $|S_r,m_r\rangle$
$S=0,1,2$, $m=-S,\ldots,S$ the spin interaction term ${\bf S}(x){\bf S}(y)$
at two neighbouring sites $x\in P_l$, $y\in P_r$ induces a change in the spin
quantum numbers
\begin{eqnarray}
|S_l,m_l\rangle\begin{array}{c}{\bf S}(x) \\ \longrightarrow\end{array}
|S'_l,m'_l\rangle & \quad &
|S_r,m_r\rangle\begin{array}{c}{\bf S}(y) \\ \longrightarrow\end{array}
|S'_r,m'_r\rangle\,. \nonumber\\
\label{spin_chge}
\end{eqnarray}
The transition matrix element
\begin{eqnarray}
\langle S'_l,m'_l|S_q(x)|S_l,m_l\rangle & = & 
v_q\left(\begin{array}{c|cc}S'_l & 1 & S_l\\m'_l & q & m_l\end{array}\right)M(S'_l,x,S_l)
\nonumber\\
 & & \label{wigner-eckart}
\end{eqnarray}
can be evaluated by means of the Wigner-Eckart Theorem.\cite{edmonds60} 
All these matrix elements
can be expressed in terms of a Clebsch-Gordan coefficient and one reduced matrix element
$M(S'_l,x,S_l)$. The latter only depends on the initial and final plaquette spins
$S_l$ and $S'_l$ and the triplet  operator $S_q(x)$ at site $x$.
The phase $v_q$ ($v_+=-1$, $v_0=v_-=1$) results from the transformation
properties of the spin operator $S_q(x)$ under the group $SU(2)$.
The interaction between neighbouring plaquettes then depends on the product of
two Clebsch-Gordan coefficients
\begin{eqnarray}
v_qv_q
\left(\begin{array}{c|cc}S'_l & 1 & S_l\\m'_l & q & m_l\end{array}\right)
\left(\begin{array}{c|cc}S'_r & 1 & S_r\\m'_r & q & m_r\end{array}\right)
\overline{M}(S'_l,S_l;S'_r,S_r) & & \nonumber\\
 & & \label{A1}
\end{eqnarray}
and the product of reduced matrix elements
\begin{eqnarray}
\overline{M}(S'_l,S_l;S'_r,S_r) & = & \sum_{\langle x,y\rangle}
M(S'_l,x,S_l)M(S'_r,y,S_r)\quad\quad\nonumber\\
\label{overline_M}
\end{eqnarray}
summed over the two neighbouring sites $\langle x,y\rangle$, which connect the left
and right plaquette as shown in Fig. \ref{fig_2}. 
The Clebsch-Gordan coefficients lead to selection rules
\begin{eqnarray}
S'_l & = & S_l\pm 1,S_l\hspace{1.25cm}\mbox{for } S_l\geq 1\nonumber\\
S'_l & = & 1 \hspace{2.2cm}\mbox{for } S_l=0\nonumber\\
m'_l & = & m_l+q\,.\nonumber\\
\label{selection_rules}
\end{eqnarray}

The explicit calculation of the transition matrix elements
\footnote{Note, that we always and without loss of correctness 
define the reduced matrix elements such that $\overline M(S_l',S_l;S_r',S_r)$
only depend on $S'_l\geq S_l$, $S'_r\geq S_r$.}yields the
weights $\overline{M}(S'_l,S_l;S'_r,S_r)$ for the following cases:

\begin{enumerate}
\item
Creation of triplet pairs from singlet pairs
\begin{eqnarray}
0\quad 0 & \leftrightarrow & A_q\quad A_{-q}
\label{CP1}
\end{eqnarray}
Note, that the initial and final states have total spin 0. According to
(\ref{A1}) and (\ref{overline_M}) the weight for this process turns out to be
\begin{eqnarray}
 & & \sum_{q}
\left(\begin{array}{c|cc} 1 & 1 & 0\\q & q & 0\end{array}\right)
\left(\begin{array}{c|cc} 1 & 1 & 0\\-q & -q & 0\end{array}\right)
\overline M^{(1)}(10,10)
\label{1010}\nonumber\\
\end{eqnarray}
where
\begin{eqnarray}
\overline M^{(1)}(10,10) & = & a^{(1)}=-\frac{1}{3}\,.
\label{W_CP1}
\end{eqnarray}

\item
Hopping of an isolated triplet on a singlet background
\begin{eqnarray}
0\quad A_q & \leftrightarrow & A_q\quad 0
\quad q=\pm 1,0
\label{0AA0}
\end{eqnarray}
Here, the initial and final states are triplets on different
plaquettes ($P_r$ and $P_l$). The corresponding weight is given again by (\ref{W_CP1}).

\item
The triplet-triplet process
\begin{eqnarray}
A_{s+q}\quad A_{t-q} & \leftrightarrow &  A_s\quad A_{t}\quad q=\pm 1,0
\label{AAAA}
\end{eqnarray}
introduces a new weight:
\begin{eqnarray}
 & & (-)^q
\left(\begin{array}{c|cc}1 & 1 & 1\\s+q & q & s\end{array}\right)
\left(\begin{array}{c|cc}1 & 1 & 1\\t-q & -q & t\end{array}\right)
\overline M^{(1)}(11,11)\,. \nonumber\\
\label{CP3b}
\end{eqnarray}
where 
\begin{eqnarray}
\overline M^{(1)}(11,11) & = & \frac{1}{4}\,.
\end{eqnarray}

\item
The process
\begin{eqnarray}
Q_r\quad 0 & \leftrightarrow & 
A_{r-q}\quad A_{q}
\label{CP4}
\end{eqnarray}
starts from an initial state with a spin 2 plaquette $Q_r$ and a spin 0
plaquette (singlet). In the final state we have two triplets $A_{r-q}$
$A_q$ coupled with an appropriate Clebsch-Gordan coefficient to form again
a spin 2 state. The coupling for (\ref{CP4}) turns out to be
\begin{eqnarray}
 & & 
\left(\begin{array}{c|cc}2 & 1 & 1\\r & q & r-q\end{array}\right)
\left(\begin{array}{c|cc} 1 & 1 & 0\\q & q & 0\end{array}\right)
\overline M(21,10)\nonumber
\end{eqnarray}
where 
\begin{eqnarray}
\overline M^{(1)}(21,10) & = & \frac{1}{2\sqrt{3}}\,.\label{wght2}
\end{eqnarray}

\item
The process
\begin{eqnarray}
Q_r\quad A_{-q} & \leftrightarrow & A_{r-q}\quad 0
\label{QAA0}
\end{eqnarray}
describes the transition from a spin 2 - spin 1 to a spin 1 spin 0
plaquette. It is accompanied with a weight
\begin{eqnarray}
 & & (-)^q
\left(\begin{array}{c|cc}2 & 1 & 1\\r & q & r-q\end{array}\right)
\left(\begin{array}{c|cc} 1 & 1 & 0\\-q & -q & 0\end{array}\right)
\overline M^{(1)}(21,10)\,.\nonumber\\
\end{eqnarray}

\item
The exchange of a spin 2 - spin 1 plaquette
\begin{eqnarray}
Q_r\quad A_{p} & \leftrightarrow & A_{r-q}\quad Q_{p+q}
\label{CP6}
\end{eqnarray}
carries a weight
\begin{eqnarray}
 & & 
\left(\begin{array}{c|cc}2 & 1 & 1\\r & q & r-q\end{array}\right)²
\left(\begin{array}{c|cc}2 & 1 & 1\\p+q & q & p\end{array}\right)²
\overline M^{(1)}(21,21)\,.\nonumber\\
\end{eqnarray}
where 
\begin{eqnarray}
\overline M^{(1)}(21,21) & = & -\frac{1}{4}\,.\label{wght3}
\end{eqnarray}

\item
If the spin 2 - spin 1 plaquettes do not change their position
\begin{eqnarray}
Q_r\quad A_{p} & \leftrightarrow & Q_{r+q}\quad A_{p-q}
\label{QAQA}
\end{eqnarray}
the corresponding weight is:
\begin{eqnarray}
 & & (-)^q
\left(\begin{array}{c|cc}2 & 1 & 2\\r+q & q & r\end{array}\right)²
\left(\begin{array}{c|cc}1 & 1 & 1\\p-q & -q & p\end{array}\right)²
\overline M^{(1)}(22,11)\,.\nonumber\\
 & & \label{W_CPf}
\end{eqnarray}
where 
\begin{eqnarray}
\overline M^{(1)}(22,11) & = & \frac{\sqrt{3}}{4}\,.
\end{eqnarray}

\end{enumerate} 

(\ref{W_CP1})-(\ref{W_CPf}) is a complete list of processes where
only singlets $|0\rangle$, $A$-triplets $|A_q\rangle$ and quintuplets
$|Q_r\rangle$ are involved. In our opinion they are most important
for the low energy behaviour for the following reasons:

\begin{itemize}
\item
$|0\rangle$ and $|A_q\rangle$ have the lowest energies according
to Table \ref{table1}.

\item
the staggered spin operator [eq. (\ref{staggered}) for $i=A$] on
the plaquette induces the dominant transitions
\begin{eqnarray}
|0\rangle\stackrel{S_q^{(*)}}{\rightarrow} |A_q\rangle
\stackrel{S_{r-q}^{(*)}}{\rightarrow} |Q_r\rangle & & \nonumber
\end{eqnarray}
As a consequence, the weights (\ref{W_CP1}), (\ref{wght2}) and
(\ref{wght3}) which are built up from these transitions are dominant
as well.
\end{itemize}

Truncating the states with subdominant transitions anticipates
antiferromagnetic order of the system for the renormalization procedure.
This is somewhat in analogy to treatments by spin-wave theory or mappings to
nonlinear sigma models, see e.g. (\onlinecite{hamer94}) and
(\onlinecite{chakravarty88}). In these approaches long-range antiferromagnetic
order is assumed from the beginning and fluctuations around this is built in
subsequently. In our approach however, the system may or may not develop
long-range order. This is determined by the renormalization group flow.


\section{The four plaquette system \label{sec3}}
\setcounter{equation}{0}

We are going to construct in this Section the ground states of the
four plaquette system with $4\times 4$ sites depicted in Fig. \ref{fig_1}.
It turns out that the ground states are symmetric under rotation of
the four plaquettes. We therefore start from rotationally symmetric
basis states, which are eigenstates of the total spin squared ${\bf S}^2$
and its 3-component
\begin{eqnarray}
{\bf S} & = & \sum_{j=1}^4{\bf S}(P_j)\,.
\end{eqnarray}



\subsection{The singlet sector}

In Table \ref{table3a} we list 7 singlet states which can be constructed
on the four plaquette system with singlets ($0$), triplets ($A_q$) and at
most one spin 2 ($Q_r$) plaquette.

\begin{table}[h!]
\begin{tabular}{cl}
\hspace{0.0cm}$|1,0\rangle =$ & $\left(\begin{array}{cc}0 & 0\\0 & 0\end{array}\right)$\\
$|2,0\rangle =$ & ${\displaystyle \frac{1}{2\sqrt{3}}\sum_q}(-)^q\left\{
\left(\begin{array}{cc}A_q & A_{-q}\\0 & 0\end{array}\right)+\mbox{ rot}\right\}$\\
$|3,0\rangle =$ & ${\displaystyle \frac{1}{\sqrt{6}}\sum_q}(-)^q\left\{
\left(\begin{array}{cc}A_q & 0\\0 & A_{-q}\end{array}\right)+
\left(\begin{array}{cc}0 & A_q\\A_{-q} & 0\end{array}\right)
\right\}$\\
$|4,0\rangle =$ & ${\displaystyle \frac{1}{3}\sum_{q,p}}(-)^{p+q}
\left(\begin{array}{cc}A_q & A_p\\A_{-p} & A_{-q}\end{array}\right)$\\
$|5',0\rangle =$ & ${\displaystyle \sum_{q,p}}(-)^{p+q}\left\{
\left(\begin{array}{cc}A_{-p} & A_p\\ A_q & A_{-q}\end{array}\right)
+\mbox{ rot}\right\}$\\\hline
$|6,0\rangle =$ & ${\displaystyle \frac{1}{2\sqrt{10}}\sum_{q,p}}(-)^{p+q}
\left(\begin{array}{c|cc}2 & 1 & 1\\r & q & p\end{array}\right)\times$\\
 & $\left\{
\left(\begin{array}{cc}A_q & A_p\\Q_{-r} & 0\end{array}\right)+
\left(\begin{array}{cc}A_q & Q_{-r}\\A_p & 0\end{array}\right)+\mbox{ rot}\right\}$\\
$|7,0\rangle =$ & ${\displaystyle \frac{1}{2\sqrt{5}}\sum_{q,p}}(-)^{p+q}
\left(\begin{array}{c|cc}2 & 1 & 1\\r & q & p\end{array}\right)\times$\\
 & $\left\{
\left(\begin{array}{cc}Q_{-r} & A_p\\A_q & 0\end{array}\right)+
\mbox{ rot}\right\}$\\
\end{tabular}
\caption{Basis states of the 4 plaquette system in the singlet sector.
}
\label{table3a}
\end{table}

Starting from the state $|1,0\rangle$, where the four plaquettes are occupied
with plaquette singlets $|0\rangle$, the creation process (\ref{CP1}) generates
the state $|2,0\rangle$ with neighbouring triplets $A_qA_{-q}$ coupled to a
total spin 0. The hopping process (\ref{0AA0}) leads from $|2,0\rangle$ to
$|3,0\rangle$. Further application of (\ref{CP1}) and (\ref{0AA0}) generates
the states $|3,0\rangle$,, $|4,0\rangle$, $|5',0\rangle$. The first four states
$|i,0\rangle$, $i=1,2,3,4$ are orthonormal. This is not the case for $|5',0\rangle$,
which is orthonormalized by
\begin{eqnarray}
|5,0\rangle & = & N_5\Big(|5',0\rangle-\langle 4,0|5',0\rangle|4,0\rangle\Big)\,,
\end{eqnarray}
where
\begin{eqnarray}
\langle 4,0|5',0\rangle & = & 4
\end{eqnarray}
and
\begin{eqnarray}
N_5 & = & \frac{1}{4\sqrt{5}}\,.
\end{eqnarray}

The states $|6,0\rangle$ and $|7,0\rangle$ contain one spin 2 plaquette $Q_r$
coupled together with two spin 1 plaquettes $A_q$ $A_p$ to form a state with
total spin 0. The states $|i,0\rangle$ $i=1,2,..,7$ are orthonormal and
complete in the sense, that no further rotational symmetric state can be
constructed from plaquette singlets, triplets and \underline{one} quintuplet.
The Hamiltonian restricted to the Hilbert space of these seven singlet states
can be written as
\begin{eqnarray}
H_0^{(2)} & = & 4E_0^{(1)}+a^{(1)}\Delta_0^{(2)}\,.
\label{H0_S}
\end{eqnarray}
The first term is just the energy of the state $|1,0\rangle$. We have scaled
out the singlet-triplet coupling (\ref{W_CP1}). The remaining ``interaction
matrix'' $\Delta_0^{(2)}$ is listed in Appendix \ref{spin0}.

The following remarkable features can be observed in the interaction
matrix $\Delta_0^{(2)}$:

\begin{enumerate}

\item
The nondiagonal matrix elements are nonnegative and fixed by the weights
\begin{eqnarray}
a^{(1)}  =  \overline M^{(1)}(10;10) & = & -\frac{1}{3}\label{a1}\\
\gamma^{(1)}  =\frac{\overline M^{(1)}(21;10)}
{\overline M^{(1)}(10;10)} & = & -\frac{\sqrt{3}}{2}
\label{gamma1}
\end{eqnarray}
They are induced by the singlet-triplet $(1,0)$ and triplet-quintuplet
$(2,1)$
transition matrix elements according to (\ref{overline_M}).

Therefore, the Perron-Frobenius theorem holds, which states that
the eigenvectors
$|\sigma^{(2)}\rangle$ with largest eigenvalue $\sigma^{(2)}$:

\begin{eqnarray}
\Delta_0^{(2)}|\sigma^{(2)}\rangle & = & \sigma^{(2)}|\sigma^{(2)}\rangle\\
\sigma^{(2)} & = & 2.20917
\end{eqnarray}
have nonnegative components:
\begin{eqnarray}
\langle i,0|\sigma^{(2)}\rangle & \geq & 0\quad i=1,...,7\,.
\end{eqnarray}

\item
The triplet-triplet and quintuplet-quintuplet transition matrix elements,
which define the weights
\begin{eqnarray}
\beta^{(1)} & = & \frac{M^{(1)}(11;11)}{a^{(1)}}=-\frac{3}{4}
\label{beta1}\\
\varepsilon^{(1)} & = & \frac{M^{(1)}(22;11)}{a^{(1)}}=
-\frac{3\sqrt{3}}{4}\label{vareps1}
\end{eqnarray}
only contribute to the diagonal matrix elements.
They also depend on the two scaled gaps
\begin{eqnarray}
\rho=\rho^{(1)} & = & \frac{1}{a^{(1)}}\Big(E_1^{(1)}-E_0^{(1)}\Big)=-3\label{rho1}\\
\kappa=\kappa^{(1)} & = & \frac{1}{a^{(1)}}\Big(E_2^{(1)}+E_0^{(1)}-
2E_1^{(1)}\Big)=-3\nonumber\\
 & & \label{kappa1}\,.
\end{eqnarray}

\item
The ground state energy of the Hamiltonian (\ref{H0_S}) in the restricted
Hilbert space of the singlet states $|j,0\rangle$, $j=1,...,7$ is given by
\begin{eqnarray}
E_0^{(2)} & = & 4E_0^{(1)}+a^{(1)}\sigma^{(2)}=-8.7236
\end{eqnarray}
which deviates from the ``exact´´ value for the $4\times 4$ system with
open b.c.
\begin{eqnarray}
E_0^{(L)} & = & -9.1892
\end{eqnarray}
by $5.0\%$.

\end{enumerate}

\subsection{The triplet sector}

\begin{table}[ht!]
\begin{tabular}{cl}
$|1,+\rangle =$ & ${\displaystyle \frac{1}{\sqrt{4}}}\left\{
\left(\begin{array}{cc}A_+ & 0\\0 & 0\end{array}\right)+\mbox{ rot}\right\}
$\\
$|2,+\rangle =$ & ${\displaystyle \frac{1}{2\sqrt{3}}\sum_q}(-)^q\left\{
\left(\begin{array}{cc}A_+ & A_q\\ A_{-q} & 0\end{array}\right)
+\mbox{ rot}\right\}$\\
$|3',+\rangle =$ & ${\displaystyle \sum_q}(-)^q\left\{
\left(\begin{array}{cc}A_+ & A_{q}\\0 & A_{-q}\end{array}\right)+
\left(\begin{array}{cc}A_q & A_{-q}\\0 & A_{+}\end{array}\right)
+\mbox{ rot}\right\}$\\\hline
$|4,+\rangle =$ & ${\displaystyle \frac{1}{2\sqrt{2}}\sum_{q}}
\left(\begin{array}{c|cc}1 & 2 & 1\\1 & 1-q & q\end{array}\right)\times$\\
 & $\hspace{1.1cm}\left\{
\left(\begin{array}{cc}Q_{1-q} & A_{q}\\0 & 0\end{array}\right)+
\left(\begin{array}{cc}Q_{1-q} & 0 \\A_{q} & 0\end{array}\right)+\mbox{ rot}
\right\}$\\
$|5,+\rangle =$ & ${\displaystyle \frac{1}{\sqrt{4}}\sum_{q}}
\left(\begin{array}{c|cc}1 & 2 & 1\\1 & 1-q & q\end{array}\right)
\left\{
\left(\begin{array}{cc}Q_{1-q} & 0\\0 & A_{q}\end{array}\right)+
\mbox{ rot}\right\}$\\
$|k,+\rangle =$ & ${\displaystyle \frac{1}{\sqrt{4}}\sum_{p,q,r}}
C_{-r,-p,-q}^{J,j}\times\left\{
\left(\begin{array}{cc}Q_{r+1} & A_{-p}\\A_{-q} & A_{-r+p+q}\end{array}
\right)+\mbox{ rot}\right\}$\\
$\mbox{with: }$ & \\
 & $C_{-r,-p,-q}^{J,j}=\left(
\begin{array}{c|cc}j & 1 & 1\\-p-q & -p & -q\end{array}\right)\times$\\
 &
$\left(\begin{array}{c|cc}J & j & 1\\-r & -p-q & -r+p+q\end{array}\right)
\left(\begin{array}{c|cc}1 & 2 & J\\1 & r+1 & -r\end{array}\right)$\\[10pt]
$\mbox{and: }$ & $\hspace{1.0cm}\begin{tabular}{c||c|ccc}
$\,k\,$ & \,6\, &  \,7 & \,8\, & 9\\ \hline
$j$ & 0 & 2 & 2 & 2\\ \hline
$J$ & 1 & 1 & 2 & 3
\end{tabular}$\\
\end{tabular}
\caption{Basis states of the 4 plaquette system in the triplet sector.}
\label{table4a}
\end{table}

The rotational symmetric eigenstates of the 4 plaquette system in the
sector with total spin $S=1$ are listed in Table \ref{table4a}:

Starting from $|1,+\rangle$, we generate the other states $|3',+\rangle$,
$|2,+\rangle$, $|4,+\rangle$ and $|5,+\rangle$ by means of the processes
(\ref{CP1}) (\ref{0AA0}) and (\ref{CP4}).
The states $|1,+\rangle$, $|2,+\rangle$, $|4,+\rangle$, $|5,+\rangle$ are
orthonormal. The state $|3',+\rangle$ is not yet orthonormal with
respect to $|2,+\rangle$, which is achieved by:
\begin{eqnarray}
|3,+\rangle & = & N_3\Big(|3',+\rangle-\langle 2,+|3',+\rangle|2,+\rangle\Big)\,,
\end{eqnarray}
with
\begin{eqnarray}
\langle 2,+|3',+\rangle & = & \frac{4}{\sqrt{3}}
\end{eqnarray}
and
\begin{eqnarray}
N_3 & = & \frac{\sqrt{3}}{4\sqrt{5}}\,.
\end{eqnarray}

The states $|k,+\rangle$, $k=6,7,8,9$ contain one spin-2 and three
spin-1 plaquettes, which are coupled together with appropriate
Clebsch-Gordan coefficients ($C_{-r,-p,-q}^{J,j}$)
eigenstates with total spin 1. There exist two further states
($j=1$, $J=1,2$) that do not couple to the considered ones.


The Hamiltonian in the restricted Hilbert space of the states $|k,+\rangle$
$k=1,\ldots,9$ reads
\begin{eqnarray}
H_1^{(2)} & = & 3E_0^{(1)}+E_1^{(1)}+a^{(1)}\Delta_1^{(2)}\,.
\label{H0_T}
\end{eqnarray}
The first term is the energy of the lowest state $|1,+\rangle$. Again we
have scaled out
the singlet-triplet coupling $a^{(1)}$ (\ref{a1}), such that the
interaction matrix $\Delta_1^{(2)}$ depends on the two scaled gaps $\rho^{(1)}$
(\ref{rho1}), $\kappa^{(1)}$ (\ref{kappa1}) and the three scaled couplings
$\gamma^{(1)}$ (\ref{gamma1}),
$\beta^{(1)}$ (\ref{beta1}) and $\varepsilon^{(1)}$ (\ref{vareps1}).
Diagonalizing the interaction matrix:
\begin{eqnarray}
\Delta_1^{(2)}|\tau^{(2)}\rangle & = & \tau^{(2)}|\tau^{(2)}\rangle
\end{eqnarray}
yields the largest eigenvalue
\begin{eqnarray}
\tau^{(2)} & = & 3.41009
\end{eqnarray}
which corresponds to a ground state energy
\begin{eqnarray}
E_1^{(2)} & = & -7-\frac{\tau^{(2)}}{3}=-8.18405\,.
\end{eqnarray}
The latter deviates from the Lanczos result on a $4\times 4$ system with open b.c.
\begin{eqnarray}
E_1^{(L)} & = & -8.6869..
\end{eqnarray}
by $6.3\%$.

\subsection{The spin 2 sector}

The 14 basis states $|l,2+\rangle$ for the 4-plaquette system
(Fig. \ref{fig_1})
in the spin 2 sector are listed in Table \ref{table5a}

The states $|l,2+\rangle$, $l=1,2,3,5$ are orthonormal,
which is not the case for $|4',2+\rangle$ with respect to $|3,2+\rangle$.
We therefore introduce
\begin{eqnarray}
|4,2+\rangle & = & N_4\Big(|4',2+\rangle-\langle 3,2+|4',2+\rangle|3,2+\rangle\Big)
\nonumber\\
\end{eqnarray}
with
\begin{eqnarray}
\langle 3,2+|4',2+\rangle & = & \frac{4\sqrt{2}}{\sqrt{3}}\\
N_4 & = & \frac{\sqrt{3}}{2\sqrt{7}}
\end{eqnarray}

The states $|6+j,2+\rangle$, $|9+j,2+\rangle$ and $|12+j,2+\rangle$
contain one spin 2 and two spin 1 plaquettes - the latter are
coupled together to a state with spin $j$, which then forms with
the $Q_r$ plaquette a state with total spin 2.

In the restricted Hilbert space of the states $|l,2+\rangle$,
$l=1,\ldots,14$
the Hamiltonian can be written as
\begin{eqnarray}
H_2^{(2)} & = & 2(E_0^{(1)}+E_1^{(1)})+a^{(1)}\Delta_2^{(2)}\,.
\label{H22}
\end{eqnarray}
Again the first term corresponds to the plaquette energies of the state
$|1,2+\rangle$ (and $|2,2+\rangle$). The interaction matrix $\Delta_2^{(2)}$
depends on the scaled gaps $\rho^{(1)}$ (\ref{rho1}), $\kappa^{(1)}$ (\ref{kappa1})
and the three scaled couplings $\gamma^{(1)}$ (\ref{gamma1}), $\beta^{(1)}$
(\ref{beta1}) and $\varepsilon^{(1)}$ (\ref{vareps1}), as can be seen in
Appendix \ref{spin2}.

\begin{table}[ht!]
\begin{tabular}{cl}
$|1,2+\rangle =$ & ${\displaystyle \frac{1}{\sqrt{4}}}\left\{
\left(\begin{array}{cc}A_+ & A_+\\0 & 0\end{array}\right)+\mbox{ rot}\right\}
$\\
$|2,2+\rangle =$ & ${\displaystyle \frac{1}{\sqrt{2}}}\left\{
\left(\begin{array}{cc}A_+ & 0\\0 & A_+\end{array}\right)+
\left(\begin{array}{cc}0 & A_+\\A_+ & 0\end{array}\right)\right\}
$\\
$|3,2+\rangle =$ & ${\displaystyle \sum_{q}\frac{(-)^q}{\sqrt{6}}}
\Bigg\{
\left(\begin{array}{cc}A_+ & A_q\\ A_{-q} & A_+\end{array}\right)+
\left(\begin{array}{cc}A_q & A_+\\ A_+ & A_{-q} \end{array}\right)
\Bigg\}$\\
$|4',2+\rangle =$ & ${\displaystyle \sum_{q}}(-)^q
\Bigg\{
\left(\begin{array}{cc}A_+ & A_+\\A_q & A_{-q}\end{array}\right)
+\mbox{ rot}\Bigg\}$\\\hline
$|5,2+\rangle =$ & ${\displaystyle \frac{1}{\sqrt{4}}}\left\{
\left(\begin{array}{cc}Q_{2+} & 0\\0 & 0\end{array}\right)+\mbox{ rot}\right\}
$\\
$|6+j,2+\rangle=$ & ${\displaystyle \frac{1}{\sqrt{4}}\sum_{q,p}}C_{p,q}^j\left\{
\left(\begin{array}{cc}A_{-p} & 0\\Q_{r+2} & A_{-q}\end{array}\right)+
\mbox{ rot}\right\}$\\
$|9+j,2+\rangle=$ & ${\displaystyle \frac{1}{\sqrt{4}}\sum_{q,p}}
C_{p,q}^j\left\{
\left(\begin{array}{cc}A_{-q} & A_{-p} \\Q_{r+2} & 0\end{array}\right)+
\mbox{ rot}\right\}$\\
$|12+j,2+\rangle=$ & ${\displaystyle \frac{(-)^j}{\sqrt{4}}\sum_{q,p}}
C_{p,q}^j\left\{
\left(\begin{array}{cc}0 & A_{-p} \\Q_{r+2} & A_{-q}\end{array}\right)+
\mbox{ rot}\right\}$\\
 & $C_{p,q}^j=
\left(\begin{array}{c|cc}2 & 2 & j\\2 & 2+r & -r\end{array}\right)
\left(\begin{array}{c|cc}j & 1 & 1\\-r & -q & -p\end{array}\right)$\\
 &  $\hspace{2.2cm}j=0,1,2;\quad r=p+q$\\
\end{tabular}
\caption{Basis states of the 4 plaquette system in the quintuplet sector.
}
\label{table5a}
\end{table}

Diagonalizing the interaction matrix
\begin{eqnarray}
\Delta_2^{(2)}|\xi^{(2)}\rangle & = & \xi^{(2)}|\xi^{(2)}\rangle
\end{eqnarray}
yields the largest eigenvalue
\begin{eqnarray}
\xi^{(2)} & = & 3.48987
\end{eqnarray}
which corresponds to a ground state energy
\begin{eqnarray}
E_2^{(2)} & = & -6-\frac{\xi^{(2)}}{3}=-7.36767\label{E0_Q}\,.
\end{eqnarray}
The latter deviates from the Lanczos result on a $4\times 4$
system with open b.c.
\begin{eqnarray}
E_2^{(L)} & = & -7.7909..
\end{eqnarray}
by $8.0\%$.

\section{The renormalization group procedure\label{sec4}}

In the previous Section we have explained how to construct the ground state
on a ($n=2$) cluster of size $4\times 4$ from four interacting $n=1$ plaquettes
($2\times 2$). We only took into account plaquette states with total spin 0, 1
and 2. This procedure will now be extended to compute the ground states on
($n+1$) clusters ($2^{n+1}\times 2^{n+1}$) with total spin $S=0,1,2$ from the
corresponding quantities on $n$ clusters. The ground states
\begin{eqnarray}
|S,m;n+1\rangle &  & S=0,1,2, m=-S,\ldots S
\label{G-S}
\end{eqnarray}
are supposed to carry alternating momenta
\begin{eqnarray}
{\bf p}_S=(0,0) & \mbox{ for }  & S=0,2\nonumber\\
{\bf p}_S=(\pi,\pi) & \mbox{ for }  & S=1\,.\nonumber\\
\end{eqnarray}
Our approach starts from the basic assumption, that the ground state (\ref{G-S})
on ($n+1$) clusters can be constructed from the ground state on $n$ clusters
\footnote{
Of course the full Hilbert space contains many more states on $n$-clusters --
like higher spin states and states with momenta ${\bf p}\ne {\bf p}_S$ like the
$B$ and $C$ triplets on the $n=1$ plaquette (Table \ref{table1}).}
\begin{eqnarray}
|S',m';n\rangle &  & S=S'+1,S',S'-1 \hspace{0.5cm}S'\geq 1\nonumber\\
 &  & S=1,0 \hspace{2.5cm}S'=0\nonumber\\
\end{eqnarray}

Under this assumption the Hamiltonians $H_S^{(n+1)}$ on the $(n+1)$ cluster
($2^{n+1}\times 2^{n+1}$) can be written in an analogous form to (\ref{H0_S})
for $S=0$, (\ref{H0_T}) for $S=1$ and (\ref{H22}) for $S=2$. To be definite,
we introduce first the analogues of the basis states in Tables \ref{table3a},
\ref{table4a}, \ref{table5a}:
\begin{eqnarray}
|i,0;n+1\rangle & \hspace{1.0cm} & i=1,\ldots 7\label{Ba}\\
|k,+;n+1\rangle & & k=1,\ldots 9\\
|l,2+;n+1\rangle & & l=1,\ldots 14\label{Bc}
\end{eqnarray}
on a $(n+1)$ block ($2^{n+1}\times 2^{n+1}$).

Then the analogues of the interaction matrices $\Delta_0^{(n+1)}$ are
obtained from Appendix \ref{appendix_a} by substituting the scaled energy gaps
and couplings
\begin{eqnarray}
\rho=\rho^{(n)} & = & \frac{E_1^{(n)}-E_0^{(n)}}{a^{(n)}}\label{scgap1}\\
\kappa=\kappa^{(n)} & = & \frac{E_2^{(n)}+E_0^{(n)}-2E_1^{(n)}}{a^{(n)}}
\label{scgap2}\\
a=a^{(n)} & = & \overline M^{(n)}(10;10)\label{coupl1}\\
\gamma=\gamma^{(n)} & = & \frac{1}{a^{(n)}}\cdot\overline M^{(n)}(21;10)
\label{coupl2}\\
\beta=\beta^{(n)} & = & \frac{1}{a^{(n)}}\cdot\overline M^{(n)}(11;11)
\label{coupl3}\\
\varepsilon=\varepsilon^{(n)} & = & \frac{1}{a^{(n)}}\cdot\overline M^{(n)}(22;11)
\label{coupl4}
\end{eqnarray}
The latter can be related by
\begin{eqnarray}
 & & \overline M^{(n)}(S_l',S_l;S_r',S_r)\nonumber\\
 & = & \sum_{\langle x,y\rangle}
M^{(n)}(S_l',x,S_l)M^{(n)}(S_r',y,S_r)\nonumber\\ \label{Mbar}
\end{eqnarray}
to the reduced matrix elements
\begin{eqnarray}
 & &\langle S_l',m_l',n|S_q(x)|S_l,m_l,n\rangle\nonumber\\
 & = & v_q
\left(\begin{array}{c|cc}S_l' & 1 & S_l\\m_l' & q & m_l\end{array}\right)
M^{(n)}(S_l',x,S_l)\nonumber\\
 & & \label{RME}
\end{eqnarray}
for the transition $S_l,m_l\rightarrow S_l',m_l'$ of the cluster spins.

The renormalization group procedure only affects the reduced matrix
elements - i.e. the couplings (\ref{coupl1}) - (\ref{coupl4}) and the
scaled gaps (\ref{scgap1}) and (\ref{scgap2}) which enter as parameters
in the analogues for the interaction matrices in Appendix \ref{appendix_a}
\begin{eqnarray}
\Delta_S^{(n+1)}(\rho^{(n)},\kappa^{(n)},\gamma^{(n)},\beta^{(n)},
\varepsilon^{(n)}) & & S=0,1,2\nonumber\\ \label{Delta}
\end{eqnarray}
on an $(n+1)$ block $2^{n+1}$.
The largest eigenvalues of the interaction matrices
\begin{eqnarray}
\Delta_0^{(n+1)}|\sigma^{(n+1)}\rangle & = & \sigma^{(n+1)}|\sigma^{(n+1)}\rangle\label{Da}\\
\Delta_1^{(n+1)}|\tau^{(n+1)}\rangle & = & \tau^{(n+1)}|\tau^{(n+1)}\rangle\\
\Delta_2^{(n+1)}|\xi^{(n+1)}\rangle & = & \xi^{(n+1)}|\xi^{(n+1)}\rangle\label{Dc}
\end{eqnarray}
yield for the ground state energies
\begin{eqnarray}
E_0^{(n+1)} & = & 4E_0^{(n)}+a^{(n)}\sigma^{(n+1)}\label{GEa}\\
E_1^{(n+1)} & = & 3E_0^{(n)}+E_1^{(n)}+a^{(n)}\tau^{(n+1)}\\
E_2^{(n+1)} & = & 2(E_0^{(n)}+E_1^{(n)})+a^{(n)}\xi^{(n+1)}\label{GEc}\,.
\end{eqnarray}

\section{The renormalization of the spin matrix elements\label{sec5}}

Our starting point is the group of spin matrix elements on a
($n+1$) cluster:
\begin{eqnarray}
\langle\tau^{(n+1)}|S_+(x)|\sigma^{(n+1)}\rangle & = & M^{(n+1)}(1,x,0)\\
\langle\xi^{(n+1)}|S_+(x)|\tau^{(n+1)}\rangle & = & M^{(n+1)}(2,x,1)\\
\langle\tau^{(n+1)}|S_0(x)|\tau^{(n+1)}\rangle & = & 
\left(\begin{array}{c|cc}1 & 1 & 1\\1 & 0 & 1\end{array}\right)
M^{(n+1)}(1,x,1)\nonumber\\
 & &\\
\langle\xi^{(n+1)}|S_0(x)|\xi^{(n+1)}\rangle & = & 
\left(\begin{array}{c|cc}2 & 1 & 2\\2 & 0 & 2\end{array}\right)
M^{(n+1)}(2,x,2)\nonumber\\
\end{eqnarray}
These matrix elements are expressed in terms of the corresponding
reduced matrix elements by means of the Wigner Eckart Theorem
[cf. (\ref{wigner-eckart})].

The eigenstates of the interaction matrices [cf. (\ref{Da})-(\ref{Dc})]
are represented in terms of the basis states [(\ref{Ba})-(\ref{Bc})]
\begin{eqnarray}
|\sigma^{(n+1)}\rangle & = & \sum_{i=1}^7\sigma_i^{(n+1)}|i,0,n+1\rangle
\label{sig_np1}\\
|\tau^{(n+1)}\rangle & = & \sum_{k=1}^{9}\tau_k^{(n+1)}|k,1,n+1\rangle\\
|\xi^{(n+1)}\rangle & = & \sum_{l=1}^{14}\xi_l^{(n+1)}|l,2,n+1\rangle
\end{eqnarray}
which leads to the following set of recursion formulas for the
reduced matrix elements
\begin{eqnarray}
M^{(n+1)}(1,x,0) & = & I^{(n+1)}(1,0)M^{(n)}(1,x,0)\nonumber\\
 & & +I^{(n+1)}(2,1)M^{(n)}(2,x,1)\nonumber\\
M^{(n+1)}(2,x,1) & = & G^{(n+1)}(1,0)M^{(n)}(1,x,0)\nonumber\\
 & & +G^{(n+1)}(2,1)M^{(n)}(2,x,1)\nonumber\\
M^{(n+1)}(1,x,1) & = & F_{\tau}^{(n+1)}(1,1)M^{(n)}(1,x,1)\nonumber\\
 & & +F_{\tau}^{(n+1)}(2,2)M^{(n)}(2,x,2)\nonumber\\
M^{(n+1)}(2,x,2) & = & F_{\xi}^{(n+1)}(1,1)M^{(n)}(1,x,1)\nonumber\\
 & & +F_{\xi}^{(n+1)}(2,2)M^{(n)}(2,x,2)\,.\nonumber\\\label{recursion}
\end{eqnarray}
The coefficients depend in a bilinear form on the components
of the eigenvectors
\begin{eqnarray}
I^{(n+1)}(a,b) & = & \sum_{k,i}\tau_k^{(n+1)}I_{k,i}(a,b)\sigma_i^{(n+1)}\label{coeff}\\
G^{(n+1)}(a,b) & = & \sum_{l,k}\xi_l^{(n+1)}G_{l,k}(a,b)\tau_k^{(n+1)}\\
F_{\tau}^{(n+1)}(a,a) & = & \sum_{k}\Big(\tau_k^{(n+1)}\Big)^2F_{\tau,k}(a,a)\\
F_{\xi}^{(n+1)}(a,a) & = & \sum_{l}\Big(\xi_l^{(n+1)}\Big)^2F_{\xi,l}(a,a)
\label{coeffb}\,.
\end{eqnarray}
Nonzero elements of the total spin combinations $(a,b)=(2,1)$ 
and $(a,a)=(2,2)$
are marked with boxes in the Tables of Appendix \ref{appendix_b}.

Note that the renormalization - i.e. the $n$-dependence of the
coefficients (\ref{coeff})-(\ref{coeffb}) - only enters via the eigenvector
components. The ``contractions'' $I_{j,i}(1,0)$, etc. are
independent of $n$ and solely determined by the spin matrix elements
between the basis states [(\ref{Ba})-(\ref{Bc})].
Therefore, they have to be calculated once and are listed in Appendix
\ref{appendix_b}.

We can check the quality of the recursion formulas $[M^{(n+1)}(1,x,0),
M^{(n+1)}(2,x,1)$ - cf. (\ref{recursion})] in the first step ($n=1$):
\begin{eqnarray}
I^{(2)}(1,0) & = & -0.64796 \\
I^{(2)}(2,1) & = & +0.03387 \\
G^{(2)}(1,0) & = & -0.66177 \\
G^{(2)}(2,1) & = & +0.02062  
\end{eqnarray}
if we compute the transition matrix elements for the staggered
spin operator $S^*(P)$ on a $2\times 2$ plaquette
\begin{eqnarray}
M^{(1)}(1,P,0) & = & -\frac{4}{\sqrt{6}}=-1.63299 \\
M^{(1)}(2,P,1) & = & \frac{4}{2\sqrt{2}}=1.41421 \\
M^{(2)}(1,P,0) & = & +0.64796\cdot\frac{4}{\sqrt{6}}
+0.03387\cdot\sqrt{2}\nonumber\\
 & = & 1.10601\label{res_a1} \\
M^{(2)}(2,P,1) & = & +0.66177\cdot\frac{4}{\sqrt{6}}
+0.02062\cdot\sqrt{2}\nonumber\\
 & = & 1.10983\label{res_a2}
\end{eqnarray}
and compare it with the Lanczos result on a $4\times 4$ system 
with open b.c.
\begin{eqnarray}
M^{(L)}(1,P,0) & = & 1.0857\label{res_b1}\\
M^{(L)}(2,P,1) & = & 1.1826\label{res_b2}
\end{eqnarray}
(\ref{res_a1}) and (\ref{res_a2}) deviate from (\ref{res_b1}) and 
(\ref{res_b2}) by $+1.9\,\%$ and $-6.2\,\%$, respectively.

\section{Recursion formulas for the scaled couplings and gaps\label{sec6}}

\begin{figure}[ht!]
\centerline{\hspace{0.0cm}\includegraphics[width=5.0cm,angle=0]{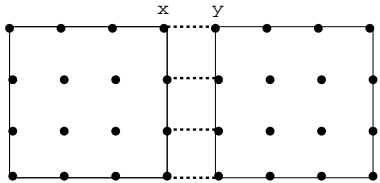}
\hspace{0.0cm}}
\caption{Couplings between neighbouring $n$-blocks -- here shown for $n=2$.
}
\label{fig_3}
\end{figure}

The relevant couplings between neighbouring $n$-blocks - as depicted in
Fig. \ref{fig_3} -
can be expressed via eqn. (\ref{Mbar}) in terms 
of the reduced matrix elements
defined through (\ref{RME}). In (\ref{Mbar}) we have to sum over 
the $2^n$ bonds 
$\langle x,y\rangle$, which connect the left and right block.
The nearest neighbour interaction ${\bf S}(x){\bf S}(y)$ on these bonds
changes the total spin on the left and right block
\begin{eqnarray}
S_l\quad\rightarrow\quad S'_l\quad & \quad &
S_r\quad\rightarrow\quad S'_r\nonumber
\end{eqnarray}
in the same way as we discussed in (\ref{spin_chge}) for the
plaquette interaction ($n=1$) as depicted in Fig. \ref{fig_2}. Note that
the definitions (\ref{Mbar}) and (\ref{overline_M}) are identical for all
blocks of sizes $2^n\times 2^n$, provided we perform the summation over
the $2^n$ connecting bonds correctly.

If we admit only blocks with total spin $0,1,2$ we get from (\ref{Mbar})
and the recursion formulas (\ref{recursion}) the renormalization of the
couplings (\ref{coupl1})-(\ref{coupl4}):

\begin{eqnarray}
\frac{a^{(n+1)}}{2a^{(n)}} & = & \Big(I^{(n+1)}(1,0)+
\gamma^{(n)}I^{(n+1)}(2,1)\Big)^2\nonumber\\
 & & \label{rec1}\\
\gamma^{(n+1)}\frac{a^{(n+1)}}{2a^{(n)}} & = & \Big(I^{(n+1)}(1,0)+
\gamma^{(n)}I^{(n+1)}(2,1)\Big)\nonumber\\
 & & \cdot\Big(G^{(n+1)}(1,0)+\gamma^{(n)}G^{(n+1)}(2,1)\Big)\nonumber\\
 & & \label{rec2}\\
\beta^{(n+1)}\frac{a^{(n+1)}}{2a^{(n)}} & = & F_{\tau}^{(n+1)}(1,1)^2\cdot\beta^{(n)}+
\nonumber\\
 & & 2F_{\tau}^{(n+1)}(1,1)F_{\tau}^{(n+1)}(2,2)\cdot\varepsilon^{(n)}
\label{rec3}\\
\varepsilon^{(n+1)}\frac{a^{(n+1)}}{2a^{(n)}} & = & F_{\tau}^{(n+1)}(1,1)F_{\xi}^{(n+1)}(1,1)
\cdot\beta^{(n)}+\nonumber\\
 & & \Big[F_{\tau}^{(n+1)}(1,1)F_{\xi}^{(n+1)}(2,2)+\nonumber\\
 & & F_{\xi}^{(n+1)}(1,1)F_{\tau}^{(n+1)}(2,2)\Big]\cdot\varepsilon^{(n)}\label{rec4}
\end{eqnarray}
In addition to the scaled couplings $\gamma^{(n)}$, $\beta^{(n)}$ and
$\varepsilon^{(n)}$ the interaction matrices (\ref{Delta}) depend on
the scaled energy differences (\ref{scgap1}), (\ref{scgap2}).
From (\ref{GEa})-(\ref{GEc}) we get the recursion formulas
\begin{eqnarray}
\frac{\rho^{(n+1)}}{\rho^{(n)}}\cdot\frac{a^{(n+1)}}{a^{(n)}} & = &
1+\fbox{J}\cdot\frac{\tau^{(n+1)}-\sigma^{(n+1)}}{\rho^{(n)}}\label{rec5}\\
\kappa^{(n+1)}\cdot\frac{a^{(n+1)}}{a^{(n)}} & = & \fbox{J}\cdot\Bigg(\xi^{(n+1)}
+\sigma^{(n+1)}-2\tau^{(n+1)}\Bigg)\label{rec6}\nonumber\\
\end{eqnarray}
Here, $\fbox{J}$ denotes the positions where a plaquette-plaquette
interaction (of general coupling strength $J$) would have to be
implemented. Throughout this work, however, we will keep $J=1$ and
discuss the interesting case of a phase transition in a two-dimensional
model of interacting plaquettes in a separate paper (\onlinecite{af08b}).

In summary, we see, that each step $n\rightarrow n+1$ in the renormalization
procedure demands the diagonalization of the three interaction matrices
$\Delta_S^{(n+1)}$ $S=0,1,2$. The eigenstates $|\sigma^{(n+1)}\rangle$,
$|\tau^{(n+1)}\rangle$, $|\xi^{(n+1)}\rangle$ with largest eigenvalues
$\sigma^{(n+1)}$, $\tau^{(n+1)}$, $\xi^{(n+1)}$ determine the right-hand
sides of the recursion formulas (\ref{rec1})-(\ref{rec6}).

\section{Numerical evaluation of the renormalization group flow\label{sec7}}

In this section we present numerical results for the evolution of
couplings and scaled gaps with $n$, which defines the block size
$2^n\times 2^n$. We start from the states in Tables \ref{table3a},
\ref{table4a}, \ref{table5a} for the singlet, triplet and spin-2
sectors. The dimensions $d_S$ of the interaction matrices $\Delta_S$
$S=0,1,2$ increases with $S$:
\begin{eqnarray}
d_0=7, \hspace{1.0cm} d_1=9, \hspace{1.0cm} d_2=14 & &
\end{eqnarray}
since the number of possibilities to construct 4 plaquette states
with singlet, triplet and at most one spin 2 plaquette increases
with $S$.

\begin{figure}[ht!]
\centerline{\hspace{0.0cm}\includegraphics[width=5.0cm,angle=-90]{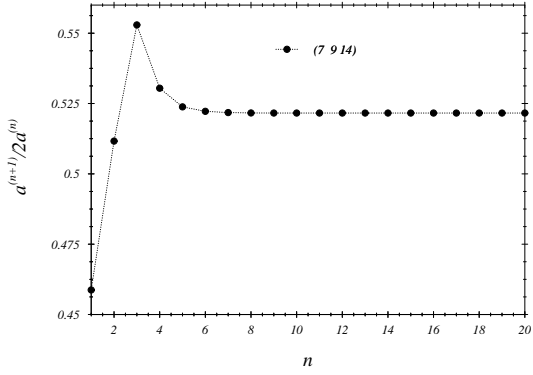}
\hspace{0.0cm}}
\caption{$n$-dependence of the coupling ratio $\frac{a^{(n+1)}}{2a^{(n)}}$
for dimensions $d_0=7$, $d_1=9$, $d_2=14$.}
\label{fig_4}
\end{figure}

In Fig. \ref{fig_4}, the ratio (\ref{rec1})
\begin{eqnarray}
\frac{a^{(n+1)}}{2a^{(n)}} & \rightarrow & \alpha<0.525
\mbox{ for }n>5
\end{eqnarray}
is shown; it converges to a constant value slightly above 1/2. 

\begin{figure}[ht!]
\centerline{\hspace{0.0cm}\includegraphics[width=5.0cm,angle=-90]{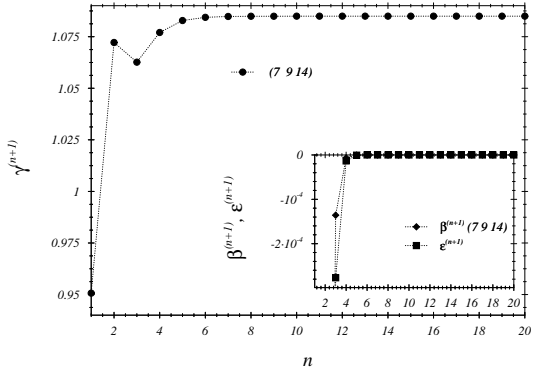}
\hspace{0.0cm}}
\caption{Evolution of the couplings $\gamma^{(n+1)}$, $\beta^{(n+1)}$, 
$\varepsilon^{(n+1)}$
for dimensions $d_0=7$, $d_1=9$, $d_2=14$.}
\label{fig_5}
\end{figure}

In Fig. \ref{fig_5}, we present the evolution of the couplings
$\gamma^{(n)}$ (\ref{rec2}), $\beta^{(n)}$ (\ref{rec3}),
$\varepsilon^{(n)}$ (\ref{rec4})
\begin{eqnarray}
\gamma^{(n)} & \rightarrow & \gamma^*=1.0847 \\
\beta^{(n)} & \rightarrow & 0\\
\varepsilon^{(n)} & \rightarrow & 0
\end{eqnarray}

This result has to be interpreted that the couplings for
\begin{itemize}
\item
the ``nondiagonal'' transitions (\ref{coupl2}), (\ref{coupl1}) with
spin exchange
\begin{eqnarray}
Q\quad A \quad  \leftrightarrow \quad  A\quad 0 &,\hspace{0.7cm}&
A\quad 0 \quad  \leftrightarrow \quad  0\quad A\nonumber
\end{eqnarray}
are both relevant in the vicinity of the fixed point.

\item
the ``diagonal'' transitions (\ref{coupl3}), (\ref{coupl4}) with no change
in the plaquette spins, however,
\begin{eqnarray}
A\quad A \quad  \leftrightarrow \quad  A\quad A &,\hspace{0.7cm}&
Q\quad A \quad  \leftrightarrow \quad  Q\quad A\nonumber
\end{eqnarray}
are irrelevant for $n\rightarrow\infty$.
\end{itemize}

\begin{figure}[ht!]
\centerline{\hspace{0.0cm}\includegraphics[width=5.0cm,angle=-90]{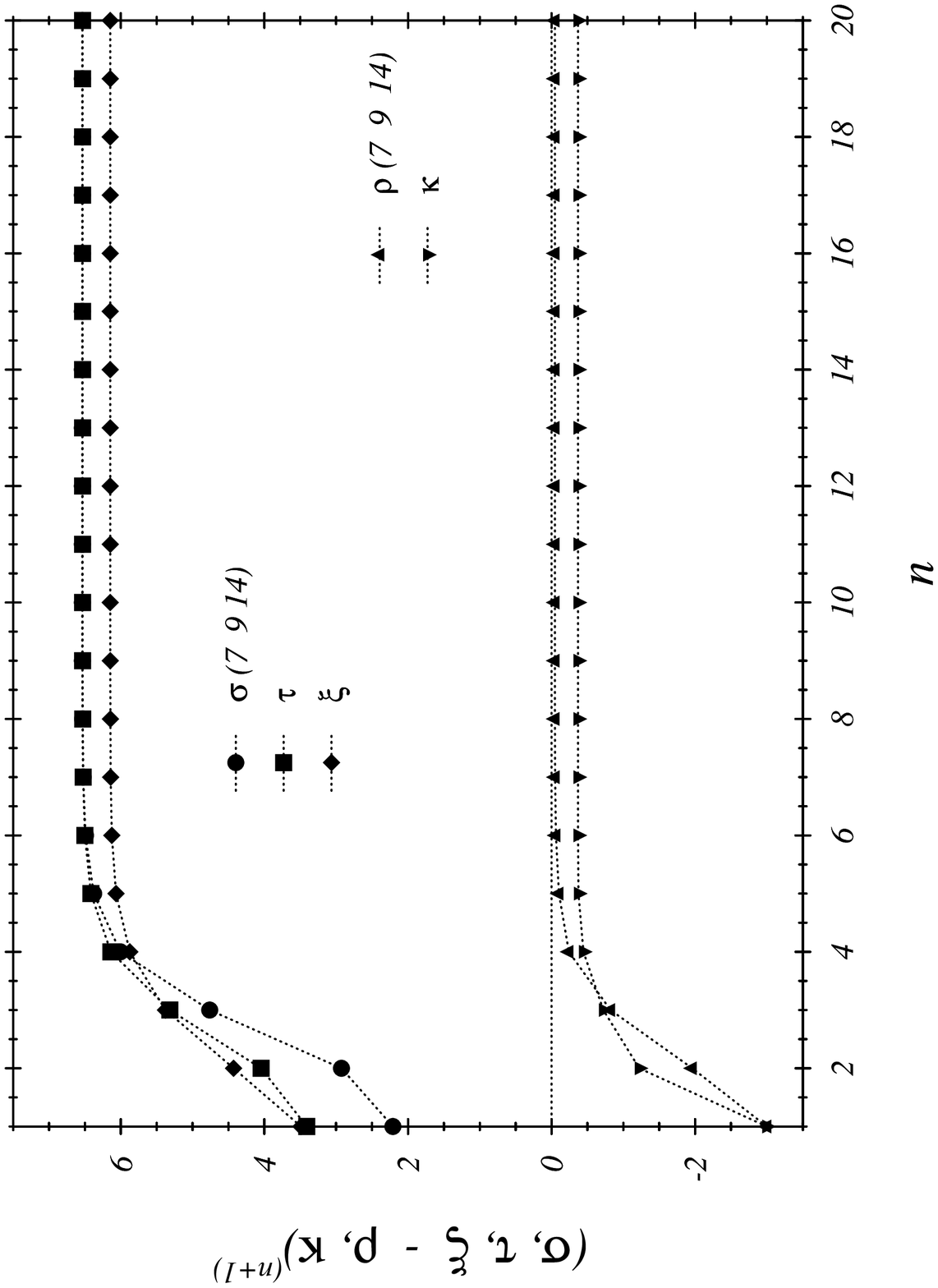}
\hspace{0.0cm}}
\caption{Evolution of the scaled energy differences $\rho^{(n)}$, $\kappa^{(n)}$
and of the largest eigenvalues $\sigma^{(n+1)}$, $\tau^{(n+1)}$, $\xi^{(n+1)}$.}
\label{fig_6}
\end{figure}

Note that the ratio $\gamma^{(n+1)}$ (\ref{rec2}) only shows a slight
variation between 1.086 and 1.151 which implies that the nondiagonal
elements in the interaction matrix $\Delta_S$, $S=0,1,2$ are almost
constant. The diagonal elements depend on the scaled energy
differences $\rho^{(n)}$ (\ref{rec5}) and $\kappa^{(n)}$ (\ref{rec6})
which increase with the system size $2^n\times 2^n$, as is shown in the
lower part of Fig. \ref{fig_6}. From the upper part we see that
the largest eigenvalues $\sigma^{(n+1)}$, $\tau^{(n+1)}$, $\xi^{(n+1)}$
of the interaction matrices increase with $n$. Indeed the essential
mechanism of the renormalization group consists in a feedback between
the increase of the (negative valued) quantities $\rho^{(n)}$,
$\kappa^{(n)}$ and the largest eigenvalues $\sigma^{(n+1)}$, 
$\tau^{(n+1)}$, $\xi^{(n+1)}$.

For large $n$
\begin{eqnarray}
\rho^{(n+1)}\rightarrow \rho^* & = & -0.046\label{rho_np1}\\ 
\kappa^{(n+1)}\rightarrow\kappa^* & = & -0.368\label{kappa_np1}\\
\sigma^{(n+1)}\rightarrow \sigma^* & = & 6.53\\
\tau^{(n+1)}\rightarrow\tau^* & = & 6.53\\
\xi^{(n+1)}\rightarrow\xi^* & = & 6.15 
\end{eqnarray}
the fixed point values $\rho^*$, $\kappa^*$ are close to zero, 
whereas $\sigma^*$, $\tau^*$, $\xi^*$ approach each other.
The deviations are a consequence of the reduced dimensions
$d_0=7$, $d_1=9$, $d_2=14$ of the Hilbert spaces for the
interaction matrices. We expect that these deviations will be
lowered, if we enlarge $d_S$, $S=0,1,2$ systematically such
that the energy differences
\begin{eqnarray}
E_1^{(n)}-E_0^{(n)} & \sim & 4^{-n\nu_1}\label{E1_0}\\
E_2^{(n)}+E_0^{(n)}-2E_1^{(n)} & \sim & 4^{-n\nu_2}\label{E2_0_1}
\end{eqnarray}
vanish in the thermodynamical limit $n\rightarrow\infty$.
The exponents
\begin{eqnarray}
\nu_1=-\frac{\log(1+x)}{\log 4} & \hspace{0.7cm}  &
\nu_2=-\frac{\log(1+y)}{\log 4} 
\end{eqnarray}
can be determined from the first derivative of the eigenvalues
$\sigma$, $\tau$, $\xi$ with respect to $\rho$ and $\kappa$,
respectively:
\begin{eqnarray}
x & = & \frac{d(\tau-\sigma)}{d\rho}\Bigg|=
\frac{\partial(\tau-\sigma)}{\partial\rho}+
\frac{\partial(\tau-\sigma)}{\partial\kappa}\cdot
\frac{d\kappa}{d\rho}\nonumber\\
 & & \label{eq_x}\\
y & = & \frac{d(\sigma+\xi-2\tau)}{d\kappa}\Bigg|=
\frac{\partial(\sigma+\xi-2\tau)}{\partial\kappa}\nonumber\\
 & & \hspace{2.7cm}+\frac{\partial(\sigma+\xi-2\tau)}{\partial\rho}\cdot
\frac{d\rho}{d\kappa}\nonumber\\
 & & \label{eq_y}
\end{eqnarray}
The partial derivatives of the eigenvalues $\sigma$, $\tau$, $\xi$ with
respect to the parameters $\rho$ and $\kappa$, which enter linearly
in the diagonals of the interaction matrices $\Delta_S(\rho,\kappa)$ 
(cf. Appendix\ref{appendix_a}) can be computed from the matrix elements of
$\frac{\partial\Delta_S}{\partial\rho}$, $\frac{\partial\Delta_S}{\partial\kappa}$,
$S=0,1,2$ between the eigenstates $|\sigma\rangle$, $|\tau\rangle$,
$|\xi\rangle$ (cf. Appendix \ref{appendix_a}).
\begin{eqnarray}
\frac{\partial\sigma}{\partial\rho} & = & \langle\sigma|\frac{\partial\Delta_0}
{\partial\rho}|\sigma\rangle=4(1-\sigma_1^2)-2(\sigma_2^2+\sigma_3^2)\nonumber\\
\frac{\partial\tau}{\partial\rho} & = & \langle\tau|\frac{\partial\Delta_1}
{\partial\rho}|\tau\rangle=4(1-\tau_1^2)-2(\tau_2^2+
\tau_3^2+\tau_4^2+\tau_5^2)\nonumber\\
\frac{\partial\xi}{\partial\rho} & = & \langle\xi|\frac{\partial\Delta_2}
{\partial\rho}|\xi\rangle=
2\cdot(1-\xi_1^2-\xi_2^2-\xi_3^2)\nonumber\\
\frac{\partial\sigma}{\partial\kappa} & = & \langle\sigma|\frac{\partial\Delta_0}
{\partial\kappa}|\sigma\rangle=\sigma_6^2+\sigma_7^2\nonumber\\
\frac{\partial\tau}{\partial\kappa} & = & \langle\tau|\frac{\partial\Delta_1}
{\partial\kappa}|\tau\rangle=1-\tau_1^2-\tau_2^2-\tau_3^2\nonumber\\
\frac{\partial\xi}{\partial\kappa} & = & \langle\xi|\frac{\partial\Delta_2}
{\partial\kappa}|\xi\rangle=1-\xi_1^2-\xi_2^2-\xi_4^2-\xi_5^2\nonumber\\
\end{eqnarray}
Remember, $\sigma_i$ ($i=1,\ldots ,7$), $\tau_k$ ($k=1,\ldots ,9$),
$\xi_l$ ($l=1,\ldots, 14$) denote the components of the eigenvectors
$|\sigma\rangle$, $|\tau\rangle$, $|\xi\rangle$, as they follow from the diagonalization of the
interaction matrices $\Delta_S(\rho,\kappa)$ for $\rho\rightarrow 0$,
$\kappa\rightarrow 0$.

\begin{figure}[ht!]
\centerline{\hspace{0.0cm}\includegraphics[width=5.0cm,angle=-90]{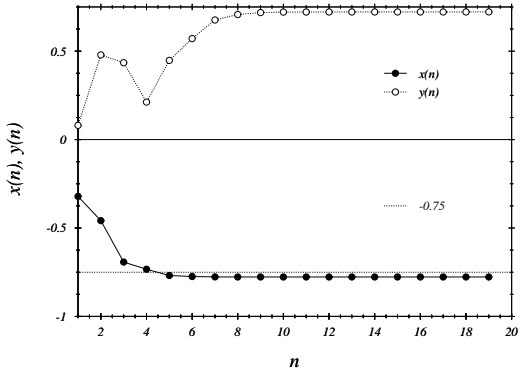}
\hspace{0.0cm}}
\caption{Quantities $x,y$ [cf. eqns. (\ref{eq_x}),(\ref{eq_y})] for
dimensions: $d_0=7$, $d_1=9$, $d_2=14$.}
\label{fig_7}
\end{figure}
In Fig. \ref{fig_7} we have plotted the quantities $x$ and $y$ vs. $n$ which
enter in (\ref{eq_x}) and (\ref{eq_y}).

The quantity $x^{(n)}$ appears to be rather stable around -0.75 for $n\geq 3$
and yields a value of $\nu_1=1$ for the critical exponent appearing in (\ref{E1_0}).
The vanishing of the singlet-triplet gap (\ref{E1_0}) in the thermodynamical
limit has been suggested by Tang and Hirsch (\onlinecite{tang89}) from a
finite-size analysis of the ground state energies.
On the other hand $y^{(n)}$ is not yet stable with respect to $n$. Again the reason
might be that our truncation of the dimensions $d_S$ for the interaction matrices
$\Delta_S$ $S=0,1,2,$ still needs to be improved.

\section{The staggered magnetization\label{sec8}}

The real space renormalization group approach generates a recursion
formula for the singlet ground state $|\sigma^{(n+1)}\rangle$ (\ref{sig_np1})
which enters in the definition of the staggered magnetization
\begin{eqnarray}
 & & \langle \sigma^{(n+1)}|\Sigma_-^{(n+1)}\Sigma_+^{(n+1)}
|\sigma^{(n+1)}\rangle\,.
\end{eqnarray}
Here
\begin{eqnarray}
\Sigma_{\pm}^{(n+1)} & = & \frac{1}{4^{n+1}}\sum_x(-)^xS_+(x)\\
 & = & \frac{1}{4}\sum_{j=1}^4\Sigma_{\pm}^{(n)}(P_j)\nonumber
\end{eqnarray}
is the properly normalized staggered spin operator on a ($n+1$)-cluster,
which can be decomposed into the corresponding quantities on the four
neighbouring $n$-clusters $P_j$, $j=1,2,3,4$, as shown in Fig. \ref{fig_1}.
This leads to the following recursion formula for the ratio
\begin{eqnarray}
R^{(n+1)} & = & \frac{\langle \sigma^{(n+1)}|\Sigma_-^{(n+1)}\Sigma_+^{(n+1)}
|\sigma^{(n+1)}\rangle}{\langle \sigma^{(n)}|\Sigma_-^{(n)}\Sigma_+^{(n)}
|\sigma^{(n)}\rangle}\nonumber\\
 & = & \sum_{i',i=1}^7\sigma_{i'}^{(n+1)}\sigma_i^{(n+1)}
\Gamma_{i',i}(\gamma^{(n)})\label{rrec_1}
\end{eqnarray}
In the evaluation of (\ref{rrec_1}) we can use the fact that the singlet
basis states $|i,0\rangle$ (Table \ref{table3a}) are invariant under
rotations of the four plaquettes $P_1,P_2,P_3,P_4$. Therefore, we are
left with the computation of the matrix elements
\begin{eqnarray}
 & & \langle i',0;n+1|\Sigma_-^{(n)}(P_j)\Sigma_+^{(n)}(P_1)|i,0;n+1\rangle,
\quad j=1,2,3\nonumber\\
\label{sc_prod}
\end{eqnarray}
which proceeds in the following steps:
\begin{itemize}
\item[a)]
The application of staggered spin operators $\Sigma_+^{(n)}(P_1)$ onto
the singlet states $|i,0;n+1\rangle$ leads to triplet states, which are
\underline{not} rotational invariant. 
A convenient set of triplet basis states is defined in Appendix
\ref{appendix_c} [(\ref{S+1_0})-(\ref{S+7_0})] together with the decomposition of
\begin{eqnarray}
 & & \Sigma_+^{(n+1)}(P_1)|i,0;n+1\rangle
\label{S+P1}
\end{eqnarray}
into these basis states [(\ref{A000})-(\ref{Q0QA})].


\item[b)]
By rotation of the 4 plaquettes
\begin{eqnarray}
 & & P_1\rightarrow P_2\quad P_2\rightarrow P_3\quad P_3\rightarrow P_4\quad
P_4\rightarrow P_1\quad\label{rotation}
\end{eqnarray}
we obtain from the decomposition of (\ref{S+P1}) the corresponding
decompositions
\begin{eqnarray}
\Sigma_+(P_2)|i,0;n+1\rangle & & \Sigma_+(P_3)|i,0;n+1\rangle
\label{decomposition_1}
\end{eqnarray}
into the triplet states (\ref{A000})-(\ref{Q0QA}).

\item[c)]
The scalar products (\ref{sc_prod}) turn out to be proportional to
\begin{eqnarray}
 & & M^{(n)}(1,P,0)^2\quad M^{(n)}(1,P,0)M^{(n)}(2,P,1)\nonumber\\
 & &  M^{(n)}(2,P,1)^2
\end{eqnarray}
where
\begin{eqnarray}
M^{(n)}(S+1,P,S) & = & \sum_{x\in P}(-)^xM^{(n)}(S+1,x,S)\nonumber\\
\end{eqnarray}
are just given by the reduced matrix elements (\ref{wigner-eckart})
summed over all sites of the $n$-cluster $P$.
\begin{eqnarray}
M^{(n)}(1,P,0)^2 & = & \langle\sigma^{(n)}|\Sigma_-^{(n)}\Sigma_+^{(n)}|
\sigma^{(n)}\rangle
\end{eqnarray}
can be identified with the staggered magnetization on the $n$-cluster $P$,
whereas the ratio
\begin{eqnarray}
\frac{M^{(n)}(2,P,1)}{M^{(n)}(1,P,0)} & = & \gamma^{(n)}
\end{eqnarray}
is given by (\ref{coupl2}).

\item[d)]
The decomposition
\begin{eqnarray}
\langle i',0;n+1|\Sigma_-^{(n)}(P_j)\Sigma_+^{(n)}(P_1)|i,0;n+1\rangle & = & 
\nonumber\\
\Gamma_{i',i}^{(j,0)}M^{(n)}(1,P,0)^2+
\Gamma_{i',i}^{(j,2)}M^{(n)}(2,P,1)^2 & & \nonumber\\
+\Gamma_{i',i}^{(j,1)}M^{(n)}(1,P,0)M^{(n)}(2,P,1)\label{decomposition_2}
\end{eqnarray}
illustrates, that the $n$-dependence - i.e. size dependence $2^{n+1}\times
2^{n+1}$ - only enters via the reduced matrix elements $M^{(n)}(1,P,0)$,
$M^{(n)}(2,P,1)$, whereas the $7\times 7$ matrices
\begin{eqnarray}
\Gamma_{i',i}^{(j,0)}, \Gamma_{i',i}^{(j,1)}, \Gamma_{i',i}^{(j,2)}, 
 &  & i,i'=1,..,7,\quad j=1,2,3\nonumber\\ 
\end{eqnarray}
are independent of $n$. They are not affected by the renormalization
group procedure and can be completely expressed in terms of scalar
products formed from the triplet states (cf. Appendix \ref{appendix_c}).

\end{itemize}

\begin{figure}[ht!]
\centerline{\hspace{0.0cm}\includegraphics[width=5.0cm,angle=-90]{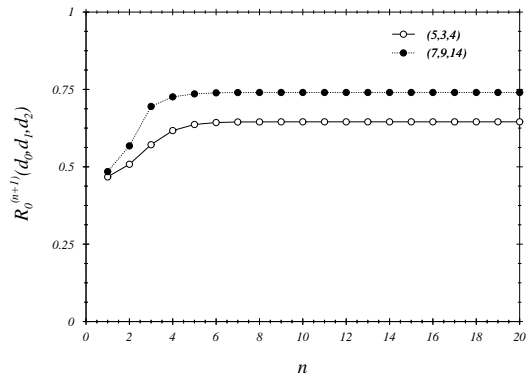}
\hspace{0.0cm}}
\caption{Evaluation of the recursion formula (\ref{rrec_1}) for
dimensions: $(d_0,d_1,d_2)=$ $(5,3,4)$, $(7,9,14)$.}
\label{fig_8}
\end{figure}

This leads to an explicit expression (\ref{Gamma1})-(\ref{Gamma4})
of the $7\times 7$ matrix $\Gamma_{i',i}(\gamma^{(n)})$, which enters
into the recursion formula (\ref{rrec_1}). The numerical evaluation
of (\ref{rrec_1}) is presented in Fig. \ref{fig_8}.

The ratio starts around $1/2$ and rapidly increases to $0.742$ and remains
constant for $n\geq 4$. A nonvanishing staggered magnetization would
demand a limiting value $R=1$ in the thermodynamical limit $n\rightarrow\infty$.
The deviation from this value, we see in Fig. \ref{fig_8}, has to be 
attributed to the truncation of the interaction matrices $\Delta_S$, $S=0,1,2$.
Their dimensions $d_S$ are limited to
\begin{eqnarray}
(d_0,d_1,d_2) & = & (7,9,14)\label{1_Quintuplet}
\end{eqnarray}
since we allow only for \underline{one} quintuplet plaquette on the four
cluster compound. We expect that the extension of the  interaction matrices
to four cluster states with $2,3,4$ quintuplets will lead to an increase of the
ratio $R$. For the moment, we can only compare the difference of including
one quintuplet (\ref{1_Quintuplet}) to zero quintuplet contributions:
\begin{eqnarray}
(d_0,d_1,d_2) & = & (5,3,4)\label{0_Quintuplet}
\end{eqnarray}
The ratio $R$ is substantially lower in the case (\ref{0_Quintuplet}) as
can be seen from Fig. \ref{fig_8}.

In other words: Higher plaquette excitations are needed to generate
plaquette-plaquette interactions which yield a nonvanishing staggered
magnetization in the thermodynamical limit.

It has been observed already by Bernu et al. (\onlinecite{bernu92}) that
the ground states in the total spin $S$ sectors collapse to the ground
state in the thermodynamical limit.

\section{Discussion and perspectives\label{sec9}}

If we compare our approach with previous renormalization group methods,
we find on one hand the same goal - namely the derivation of a low
energy effective Hamiltonian - but also crucial differences in the
underlying assumptions:

Most of the ``older'' approaches like that of Lepetit and Manousakis
(\onlinecite{manousakis93})
start with blocks with an odd number of sites. Here, the block ground
state has spin $1/2$ and the Wigner-Eckart Theorem allows already
interactions between neighbouring blocks in the ground state.
Excited states - e.g. with total spin $3/2$ - are assumed to contribute
only to a renormalization of the coupling between blocks in the ground
state. Therefore, there is no renormalization of the energy difference
between the ground state and excited states. In our opinion this is the
reason, why these approaches do not allow for spin-spin correlations
at large distances.
The exact RG flow acts in an infinite-dimensional space of Hamiltonians
resp. couplings. Even when starting with a model that is defined by very few
couplings, the exact flow will carry the Hamiltonian into rather complicated
regimes: at each step of an RG procedure longer-ranged couplings are
generated. In the past, in many applications of the RG concept the space of
all Hamiltonians was truncated to a finite dimensional one, i.e. only a few
coupling constants were kept. For many universal properties, this approach was
successful.

Recent approaches like ours and that of Capponi et al. (\onlinecite{capponi04}),
Albuquerque et al. (\onlinecite{troyer08})
based on the CORE method (Contractor Renormalization group) (\onlinecite{
morningstar94})
start with plaquettes having a singlet ground state. They cannot interact,
since the interaction is mediated by the spin operators in the Hamiltonian.
Their expectation values between total spin 0 states vanish according to
the Wigner-Eckart Theorem. Therefore excited states on the plaquettes are
absolutely necessary to generate interactions between the plaquettes [cf.
the processes (\ref{CP1}),(\ref{0AA0}),(\ref{AAAA}),(\ref{CP4}),(\ref{QAA0}),
(\ref{CP6}),(\ref{QAQA}) of Section II].
For this reason we included triplet ($|A_q\rangle$, $q=\pm 1,0$) and quintuplet
($|Q_r\rangle$, $r=\pm 2,\pm 1,0$) excitations. Indeed it turned out that the
states with one quintuplet excitation (i.e. $|6,0\rangle$ and
$|7,0\rangle$ in Table \ref{table3a} and $|k,+\rangle$, $k=4,5,..,9$
in Table \ref{table4a} and 
$|3,2+\rangle$, $|6+j,2+\rangle$, $|9+j,2+\rangle$, $|12+j,2+\rangle$ ($j=0,1,2$)
in Table \ref{table5a}) improve the decrease of the singlet-triplet gap
in the large-$n$ limit
\begin{eqnarray}
\rho^*(5,3,4)=-0.767 & ; & \rho^*(7,9,14)=-0.183
\label{stg_limit}
\end{eqnarray}
The limiting value (\ref{stg_limit}) defines a measure for the ``quality''
of the singlet-triplet gap generated with interaction matrices $\Delta_S$
of dimensions $d_S$, $S=0,1,2$
\begin{eqnarray}
(d_0,d_1,d_2)=(5,\,3,\,4) & ; & (7,\,9,\,14)
\label{stg_limit_dimension}
\end{eqnarray}
Such an extension of the interaction matrices also leads to an improvement
of the staggered magnetization, as discussed in Fig. \ref{fig_8}.

In refs. (\onlinecite{capponi04}), (\onlinecite{troyer08}) the quintuplet 
excitations are missing and
it would be interesting to see how the singlet-triplet gap decreases in
their renormalization process.
Note however, that quintuplet excitations possess large couplings (\ref{wght2})
to triplet excitations, which do not die out in the renormalization process
($n\rightarrow\infty$) (Fig. \ref{fig_5}); the corresponding energy differences
(\ref{scgap1}), (\ref{scgap2}) decrease as well (Fig. \ref{fig_6}). The authors of ref.
(\onlinecite{troyer08}) intend to improve the CORE results by varying the
compounds of plaquettes. In addition to the 4 plaquette compound of Fig.
\ref{fig_1}, they allow 2 and 3 plaquette compounds. We do not have this
freedom, since our renormalization approach is restricted to the geometry
of the 4 plaquette compound.
The restriction to rotational symmetric states on the 4 plaquette compound
with singlets, triplets and quintuplets enables us to follow the renormalization
group flow for all couplings and gaps, which enter into the interaction 
matrices.

We intend to improve our results in a first step by taking into
account all rotational symmetric states on the 4 plaquette compound
with $n_Q=2,3,4$ quintuplets. Larger interaction matrices demand
for an efficient method to calculate matrix elements which form
$SU(2)$ invariants - similar to Racah coefficients (\onlinecite{edmonds60})
in Nuclear Physics.



\vspace{2.0cm}







\begin{appendix}
\section{The interaction matrices}
\label{appendix_a}

Here we present the interaction matrices $\Delta_S$, $S=0,1,2$. In order
to clarify the dependence on the scaled energy gaps $\rho$, $\kappa$ and
coupling constants $\gamma$, $\beta$, $\varepsilon$ [cf. eqs.
(\ref{scgap1})-(\ref{coupl4})],
it is convenient to consider the following block
forms:

\subsection{Spin 0 sector\label{spin0}}

\begin{eqnarray}
\Delta_0 & = & \left(\begin{array}{c|c}
\Delta(5,5) & \Delta(5,6)\\
\Delta^T(5,6) & \Delta(6,6)
\end{array}\right)
\end{eqnarray}
$\Delta(5,5)$ is fixed by the matrix elements of the first five states:
\begin{eqnarray}
\Delta(5,5) & = &
\left(\begin{array}{c|c|c|c|c}
0 & 2\sqrt{3} & 0 & 0 & 0 \\ \hline
2\sqrt{3} & 2\rho-\beta & 2\sqrt{2} & \frac{2}{\sqrt{3}} &
2\sqrt{\frac{5}{3}} \\ \hline
0 & 2\sqrt{2} & 2\rho  & 0 & 0 \\ \hline
0 & \frac{2}{\sqrt{3}} & 0 & 4\rho  & 0 \\ \hline
0 & 2\sqrt{\frac{5}{3}} & 0 & 0 &
4\rho+\frac{1}{2}\beta
\end{array}\right)\nonumber\\
\end{eqnarray}

Note that this block matrix only depends on $\rho$ and $\beta$.

The matrix elements between the states $i=1,\ldots,5$ and $i=6,7$
are contained in the $2\times 5$ block matrix
\begin{eqnarray}
\Delta^T(5,6) & = & \gamma\left(\begin{array}{ccccc}
0 & 0 & 2\sqrt{\frac{5}{3}} & \frac{2\sqrt{10}}{3} & \frac{\sqrt{2}}{3}\\
0 & 2\sqrt{\frac{5}{3}} & 0 & 0 & 0
\end{array}\right)
\end{eqnarray}
which is proportional to $\gamma$.

The matrix elements $i=6,7$, $i'=6,7$ form the third block matrix
\begin{eqnarray}
\Delta(6,6) & = & (4\rho+\kappa)\mathbf{1}-\frac{1}{6}\beta\left(
\begin{array}{cc}1 & 0\\0 & 0
\end{array}\right)-\frac{\sqrt{3}}{2}\varepsilon\left(\begin{array}{cc}1 & 0\\
0 & 2
\end{array}\right)\nonumber\\
 & & +\Big(\sqrt{2}+\frac{1}{2\sqrt{3}}\gamma^2\Big)\left(
\begin{array}{cc}0 & 1\\1 & 0
\end{array}\right)
\end{eqnarray}

\subsection{Spin 1 sector\label{spin1}}

\begin{eqnarray}
\Delta_1 & = & \left(\begin{array}{cc}
\Delta(5,5) & \Delta(5,6)\\
\Delta^T(5,6) & \Delta(6,6)
\end{array}\right)
\end{eqnarray}
$\Delta(5,5)$ is fixed by the matrix elements of the first five states:

\begin{eqnarray}
 & & \Delta(5,5)= \nonumber \\
 & & \left(\begin{array}{c|c|c||c|c}
2 & \frac{2}{\sqrt{3}} & 2\sqrt{\frac{5}{3}} & -\gamma\sqrt{\frac{10}{3}}
 & 0 \\ \hline
\frac{2}{\sqrt{3}} & 2\rho+\frac{2}{3} & \frac{2}{3}\sqrt{5} &
-\frac{\sqrt{10}}{3}\gamma & -\frac{\sqrt{20}}{3}\gamma \\ \hline
2\sqrt{\frac{5}{3}} & \frac{2}{3}\sqrt{5} &
2\rho+\frac{1}{3}-\frac{3}{2}\beta & -\frac{1}{3\sqrt{2}}\gamma &
-\frac{1}{3}\gamma \\ \hline\hline
-\gamma\sqrt{\frac{10}{3}} & -\frac{\sqrt{10}}{3}\gamma &
-\frac{1}{3\sqrt{2}}\gamma & \kappa+2\rho+ &
\sqrt{2} \\
 & & & \frac{1}{6}\gamma^2-\frac{\sqrt{3}}{2}\varepsilon & \\\hline
 0 & -\frac{\sqrt{20}}{3}\gamma & -\frac{1}{3}\gamma & \sqrt{2} &
\kappa+2\rho
\end{array}\right)\nonumber\\
\end{eqnarray}

The matrix elements between the states $k=1,\ldots,5$ and $k=6,7,8,9$
are contained in the $4\times 5$ block matrix
\begin{eqnarray}
\Delta(5,6) & = & \left(\begin{array}{cccc}
 0 & 0 & 0 & 0 \\[5pt]
 \frac{2\sqrt{5}}{3\sqrt{3}}\gamma & \frac{1}{3\sqrt{3}}\gamma &
\frac{\sqrt{5}}{3}\gamma & \frac{2\sqrt{7}}{3}\gamma \\[5pt]
 \frac{1}{3\sqrt{3}}\gamma & \frac{23}{3\sqrt{15}}\gamma &
-\frac{4}{3}\gamma & \frac{\sqrt{7}}{3\sqrt{5}}\gamma \\[5pt]
 -\frac{\sqrt{2}}{\sqrt{3}} & -\frac{\sqrt{10}}{\sqrt{3}} & 0 & 0 \\[5pt]
 0 & 0 & 0 & 0 \\
\end{array}\right)
\end{eqnarray}

The matrix elements $k=6,7,8,9$, $k'=6,7,8,9$ form the third block matrix:
\begin{eqnarray}
\Delta(6,6) & = & (4\rho+\kappa)\cdot\mathbf{1}+
\frac{\beta}{2}\cdot\left(\begin{array}{cccc}
0 & 0 & 0 & 0\\
0 & -3 & 0 & 0\\
0 & 0 & -1 & 0\\
0 & 0 & 0 & +2\end{array}\right)\nonumber\\
\end{eqnarray}
\begin{eqnarray}
 & & +\varepsilon\cdot\left(\begin{array}{cccc}
0 & 0 & 0 & 0\\
0 & -\frac{3\sqrt{3}}{4} & +\frac{3}{4\sqrt{5}} & 0\\
0 & +\frac{3}{4\sqrt{5}} & -\frac{25\sqrt{3}}{36} &
+\frac{7\sqrt{3}}{9\sqrt{35}}\\
0 & 0 & +\frac{7\sqrt{3}}{9\sqrt{35}} & -\frac{8\sqrt{3}}{9}
\end{array}\right)\nonumber\\
 & & +\gamma^2\cdot\left(\begin{array}{cccc}
 \frac{10}{9} & \frac{\sqrt{5}}{90} & \frac{1}{6\sqrt{3}} &
\frac{\sqrt{35}}{15\sqrt{3}}  \\[5pt]
 \frac{\sqrt{5}}{90} & \frac{91}{180} & \frac{95\sqrt{3}}{180\sqrt{5}}  &
\frac{\sqrt{21}}{90}  \\[5pt]
 \frac{1}{6\sqrt{3}} & \frac{95\sqrt{3}}{180\sqrt{5}} &
-\frac{1}{12} & -\frac{\sqrt{7}}{6\sqrt{5}}  \\[5pt]
\frac{\sqrt{35}}{15\sqrt{3}}  & \frac{\sqrt{21}}{90}
  & -\frac{\sqrt{7}}{6\sqrt{5}}  & \frac{2}{15} \\
\end{array}\right)\nonumber\\
\end{eqnarray}

\subsection{Spin 2 sector\label{spin2}}

\begin{eqnarray}
\Delta_2 & = & \left(\begin{array}{c|ccc}
\Delta(5,5) & \Delta(5,6) & \Delta(5,9) &\Delta(5,12)\\\hline
\Delta^T(5,6) & \Delta(6,6) & \Delta(6,9) & \Delta(6,12) \\
\Delta^T(5,9) & \Delta^T(6,9) & \Delta(9,9) & 0\\
\Delta^T(5,12) & \Delta^T(6,12) & 0 & \Delta(12,12)
\end{array}\right)\nonumber\\
\end{eqnarray}
$\Delta(5,5)$ is fixed by the matrix elements of the first five states:
\begin{eqnarray}
\Delta(5,5) & = &
\left(\begin{array}{c|c|c|c||c}
+\frac{1}{2}\beta & 2\sqrt{2} & \frac{2\sqrt{2}}{\sqrt{3}} &
\frac{\sqrt{7}}{\sqrt{3}} & 2\gamma \\ \hline
2\sqrt{2} & 0 & 0 & 0 & 0 \\ \hline
\frac{2\sqrt{2}}{\sqrt{3}} & 0 & 2\rho & 0 & 0 \\ \hline
\frac{\sqrt{7}}{\sqrt{3}} & 0 & 0 & 2\rho+\frac{3}{7}\beta & 0\\\hline\hline
2\gamma & 0 & 0 & 0 & \kappa  
\end{array}\right)\nonumber \\
 & & \\
\Delta(5,6) & = & \gamma\cdot\left(\begin{array}{c|c|c}
-\frac{2}{\sqrt{3}} & 0 & -\frac{\sqrt{7}}{\sqrt{3}}
 \\ \hline
0 & 0 & 0 \\ \hline
0 & 0 & 0 \\ \hline
0 & 0 & 0 \\ \hline
0 & 0 & 0
\end{array}\right)\\
\Delta(5,9) & = & \left(\begin{array}{c|c|c}
0 & 0 & 0 \\ \hline
-\sqrt{\frac{2}{3}}\gamma & +\sqrt{\frac{3}{2}}\gamma &
-\frac{\sqrt{7}}{\sqrt{6}}\gamma\\ \hline
-\frac{2\sqrt{2}}{3}\gamma & 0 &
-\frac{\sqrt{2\cdot 7}}{3}\gamma \\\hline
-\frac{\sqrt{7}}{3}\gamma & 0 &
\frac{1}{3}\gamma\\\hline
-\sqrt{3} & 0 & 0
\end{array}\right)
\end{eqnarray}
\begin{eqnarray}
\Delta(5,12) & = & \left(\begin{array}{c|c|c}
0 & 0 & 0 \\ \hline
-\sqrt{\frac{2}{3}}\gamma & -\sqrt{\frac{3}{2}}\gamma &
-\frac{\sqrt{7}}{\sqrt{6}}\gamma\\ \hline
-\frac{2\sqrt{2}}{3}\gamma & 0 &
-\frac{\sqrt{2\cdot 7}}{3}\gamma \\\hline
-\frac{\sqrt{7}}{3}\gamma & 0 &
\frac{1}{3}\gamma\\\hline
-\sqrt{3} & 0 & 0
\end{array}\right)
\end{eqnarray}
\begin{eqnarray}
\Delta(6,6) & = & \Big(2\rho+\kappa\Big)\cdot\mathbf{1}+
\varepsilon\Big(K+\tilde K\Big)\\
 & & \nonumber
\end{eqnarray}
\begin{eqnarray}
K & = & \frac{1}{4\sqrt{3}}\left(\begin{array}{ccc}
0 & 4 & 0 \\
4 & -1 & \sqrt{7}\\
0 & \sqrt{7} & -3
\end{array}\right)\\
\tilde K & = & FKF;\quad F=\left(\begin{array}{ccc}
1 & 0 & 0 \\
0 & -1 & 0\\
0 & 0 & 1
\end{array}\right)
 \end{eqnarray}
 \begin{eqnarray}
\Delta(9,9) & = & \Big(2\rho+\kappa\Big)\cdot\mathbf{1}+
\varepsilon\cdot K+
\frac{\beta}{2}\left(\begin{array}{ccc}
-2 & 0 & 0 \\
0 & -1 & 0\\
0 & 0 & 1
\end{array}\right)\nonumber\\
 & & \\
\Delta(12,12) & = & \Big(2\rho+\kappa\Big)\cdot\mathbf{1}+
\varepsilon\cdot \tilde K+
\frac{\beta}{2}\left(\begin{array}{ccc}
-2 & 0 & 0 \\
0 & -1 & 0\\
0 & 0 & 1
\end{array}\right)\nonumber\\
\end{eqnarray}
\begin{eqnarray}
\Delta(6,9) & = & F+\gamma^2 C\\
\Delta(6,12) & = & F+\gamma^2 FCF
\end{eqnarray}
\begin{eqnarray}
C & = & \frac{1}{6}\left(\begin{array}{ccc}
2 & 3 & \sqrt{7} \\
3 & 3 & 0\\
\sqrt{7} & 0 & -1
\end{array}\right)
\end{eqnarray}

\section{Coefficients (\ref{coeff})-(\ref{coeffb})}
\label{appendix_b}

In this Appendix the coefficients (\ref{coeff})-(\ref{coeffb}) of Section \ref{sec5} are
given. Note that the dimensions $d_S=7,9,14$ have been used.

Note further, that those nonzero coefficients with spin 2 plaquette
contribution are marked with boxes. I.e., the reduced Hilbert space
[$(d_0,d_1,d_2)=(5,3,4)$] for zero spin 2 plaquettes is directly
illustrated.

\begin{widetext}
\begin{eqnarray}
I_{k,i} & = & \left(\begin{array}{ccccc|cc}
-\frac{1}{2} & -\frac{1}{2\sqrt{3}} & -\frac{1}{2\sqrt{6}} & 0 & 0 &
0 & 0 \\
0 & -\frac{1}{6} & -\frac{1}{2\sqrt{2}} & -\frac{1}{2\sqrt{3}} & 0 &
\fbox{$-\frac{5}{6\sqrt{30}}$} & 0\\
0 & -\frac{\sqrt{5}}{6} & 0 & 0 & -\frac{7}{20\sqrt{3}} &
\fbox{$-\frac{1}{12\sqrt{6}}$} & \fbox{$-\frac{1}{4\sqrt{3}}$}\\\hline
0 & \fbox{$-\frac{\sqrt{5}}{6\sqrt{2}}$} & 0 & 0 & 0 &
+\frac{1}{10\sqrt{3}} & +\frac{1}{5\sqrt{6}}\\
0 & 0 & \fbox{$+\frac{\sqrt{10}}{12}$} & 0 & 0 &
+\frac{\sqrt{2}}{4\sqrt{3}} & 0\\
0 & 0 & 0 & \fbox{$-\frac{16+\sqrt{6}}{35\sqrt{5}}$} & 0 &
-\frac{1}{6\sqrt{2}} & 0\\
0 & 0 & 0 & \fbox{$-\frac{22+\sqrt{6}}{360}$} &
\fbox{$-\frac{45+\sqrt{2}}{60\sqrt{5}}$} &
-\frac{1}{12\sqrt{10}} &
-\frac{7+3\sqrt{2}}{40\sqrt{5}}\\
0 & 0 & 0 & \fbox{$-\frac{10\sqrt{3}-3\sqrt{2}}{72\sqrt{5}}$} &
\fbox{$-\frac{\sqrt{3}}{60}$} &
-\frac{1}{4\sqrt{6}} & +\frac{7\sqrt{3}+3\sqrt{6}}{120}\\
0 & 0 & 0 & \fbox{$-\frac{2\sqrt{3}+3\sqrt{2}}{90\sqrt{7}}$} &
\fbox{$-\frac{2}{5\sqrt{105}}$} &
-\frac{2}{\sqrt{210}} & -\frac{3+16\sqrt{2}}{10\sqrt{210}}\\
\end{array}\right)\nonumber\\
\end{eqnarray}
\begin{eqnarray}
G_{l,k} & = & \left(\begin{array}{ccc|cccccc}
-\frac{1}{2} & -\frac{1}{2\sqrt{3}} & -\frac{1}{4\sqrt{15}} &
\fbox{$+\frac{1}{4\sqrt{30}}$} & 0 &0 & 0 & 0 & 0\\
-\frac{1}{2\sqrt{2}} & 0 & -\frac{\sqrt{6}}{4\sqrt{5}} & 0 &
\fbox{$+\frac{1}{4\sqrt{30}}$} & 0 & 0 & 0 & 0\\
0 & -\frac{1}{2\sqrt{2}} & -\frac{1}{\sqrt{10}} & 0 & 0 &
\fbox{$-\frac{1}{4\sqrt{30}}$} & \fbox{$-\frac{\sqrt{6}}{120}$} &
\fbox{$-\frac{1+2\sqrt{2}}{4\sqrt{10}}$} &
\fbox{$-\frac{\sqrt{7}}{10\sqrt{2}}$}\\
0 &  +\frac{1}{4\sqrt{7}} &
-\frac{1}{4\sqrt{5\cdot 7}} & 0 & 0 &
\fbox{$+\frac{1}{12\sqrt{105}}$} & \fbox{$-\frac{2}{3\sqrt{21}}$} &
\fbox{$+\frac{3\sqrt{5}}{12\sqrt{7}}$} &
\fbox{$-\frac{1}{84}$}\\ \hline
\fbox{$-\frac{1}{4}$} & 0 & 0 & +\frac{\sqrt{6}}{4\sqrt{5}} &
+\frac{\sqrt{3}}{4\sqrt{5}} &
0 & 0 & 0 & 0\\
0 & \fbox{$+\frac{1}{4}$} & 0 & -\frac{1}{2\sqrt{10}} & 0 &
+\frac{3}{2\sqrt{60}} & 0 & 0 & 0\\
0 & 0 & 0 & 0 & 0 & 0 & 0 & 0 & 0\\
0 & 0 & \fbox{$+\frac{\sqrt{7}}{8\sqrt{5}}$} & -\frac{\sqrt{7}}{4\sqrt{10}}
 & 0 & 0 & +\frac{11\sqrt{3}}{80\sqrt{7}} & +\frac{33}{48\sqrt{35}} &
+\frac{11}{280}\\
0 & \fbox{$+\frac{1}{12}$} & \fbox{$+\frac{1}{6\sqrt{5}}$} &
-\frac{1}{4\sqrt{10}} & -\frac{1}{4\sqrt{5}} &
+\frac{1}{4\sqrt{15}} & +\frac{1}{4\sqrt{3}} & 0 & 0\\
0 & \fbox{$+\frac{1}{8}$} & \fbox{$+\frac{1}{16\sqrt{5}}$} &
-\frac{1}{8\sqrt{10}} & -\frac{3}{8\sqrt{5}} &
+\frac{3}{8\sqrt{15}} & -\frac{\sqrt{3}}{16} &
+\frac{3}{16\sqrt{5}} & 0\\
0 & \fbox{$+\frac{\sqrt{7}}{24}$} & \fbox{$+\frac{\sqrt{7}}{48\sqrt{5}}$} &
-\frac{\sqrt{7}}{8\sqrt{10}} & -\frac{\sqrt{7}}{8\sqrt{5}} &
+\frac{\sqrt{7}}{8\sqrt{15}} & +\frac{19-12\sqrt{2}}{80\sqrt{21}} &
-\frac{17+4\sqrt{2}}{48\sqrt{35}} & +\frac{17+4\sqrt{2}}{420}\\
0 & \fbox{$+\frac{1}{12}$} & \fbox{$+\frac{1}{6\sqrt{5}}$} &
-\frac{1}{4\sqrt{10}} & -\frac{1}{4\sqrt{5}} &
+\frac{1}{4\sqrt{15}} & +\frac{1}{4\sqrt{3}} & 0 & 0\\
0 & \fbox{$-\frac{1}{8}$} & \fbox{$-\frac{1}{16\sqrt{5}}$} &
+\frac{1}{8\sqrt{10}} & +\frac{3}{8\sqrt{5}} &
-\frac{3}{8\sqrt{15}} & +\frac{\sqrt{3}}{16} &
-\frac{3}{16\sqrt{5}} & 0\\
0 & \fbox{$+\frac{\sqrt{7}}{24}$} & \fbox{$+\frac{\sqrt{7}}{48\sqrt{5}}$} &
-\frac{\sqrt{7}}{8\sqrt{10}} & -\frac{\sqrt{7}}{8\sqrt{5}} &
+\frac{\sqrt{7}}{8\sqrt{15}} & +\frac{19-12\sqrt{2}}{80\sqrt{21}} &
-\frac{17+4\sqrt{2}}{48\sqrt{35}} & +\frac{17+4\sqrt{2}}{420}\\
\end{array}\right)\nonumber\\
\end{eqnarray}
\end{widetext}

\begin{eqnarray}
F_{\xi,\xi}: &  & \begin{array}{c|cc}
j & F_{j,j} & \fbox{$F_{j,j}$} \\\hline
1 & \frac{1}{2\sqrt{2}} & -  \\ 
2 & \frac{1}{2\sqrt{2}} & - \\
3 & \frac{1}{2\sqrt{2}} & - \\
4 & -\frac{\sqrt{2}}{28} & - \\\hline\hline
5 & - & \frac{1}{2\sqrt{6}}  \\
6 & - & \frac{1}{2\sqrt{6}} \\
7 & \frac{1}{12\sqrt{2}} & \frac{5}{12\sqrt{6}} \\
8 & \frac{1}{4\sqrt{2}} & \frac{1}{4\sqrt{6}}\\
9 & - & \frac{1}{2\sqrt{6}}\\
10 & \frac{1}{12\sqrt{2}} & \frac{5}{12\sqrt{6}}\\
11 & \frac{1}{4\sqrt{2}} & \frac{1}{4\sqrt{6}}\\
12 & - & \frac{1}{2\sqrt{6}}\\
13 & \frac{1}{12\sqrt{2}} & \frac{5}{12\sqrt{6}}\\
14 & \frac{1}{4\sqrt{2}} & \frac{1}{4\sqrt{6}}
\\
\end{array}\nonumber\\
\end{eqnarray}

\begin{eqnarray}
F_{\tau,\tau}: &  & \begin{array}{c|cc}
j & F_{j,j} & \fbox{$F_{j,j}$}\\ \hline
1 & \frac{1}{4\sqrt{2}} & - \\
2 & \frac{1}{4\sqrt{2}} & - \\
3 & \frac{7}{40\sqrt{2}} & - \\\hline\hline
4 & -\frac{1}{8\sqrt{2}} & \frac{3}{8\sqrt{6}}\\
5 & -\frac{1}{8\sqrt{2}} & \frac{3}{8\sqrt{6}}\\
6 & -\frac{11}{25}\cdot\frac{1}{4\sqrt{2}} &
+\frac{3}{2}\cdot\frac{1}{4\sqrt{6}} \\
7 & -\frac{1}{2}\cdot\frac{1}{4\sqrt{2}} &
+\frac{69}{50}\cdot\frac{1}{4\sqrt{6}}\\
8 & +\frac{1}{2}\cdot\frac{1}{4\sqrt{2}} &
+\frac{1}{2}\cdot\frac{1}{4\sqrt{6}}\\
9 & +2\cdot\frac{1}{4\sqrt{2}} &
-\frac{1}{4\sqrt{6}}\\
\end{array}\nonumber\\
\end{eqnarray}


\section{Recursion formula for the staggered magnetization}
\label{appendix_c}

As explained in Section \ref{sec8} the derivation of (\ref{rrec_1})
follows the steps

\begin{itemize}
\item[a)]
The application of the staggered spin operator $\Sigma_+^{(n)}(P_1)$ onto
the singlet basis states $|i,0;n+1\rangle$ - listed in Table \ref{table3a} -
generate triplet states, which are not rotational invariant:

\begin{eqnarray}
\Sigma_+^{(n)}(P_1)|1,0\rangle  & = & 
-M^{(n)}(1,P,0)|A_+,0,0,0\rangle\label{S+1_0}\\
\Sigma_+^{(n)}(P_1)|2,0\rangle  & = &
-\frac{1}{\sqrt{12}}
M^{(n)}(1,P,0)\times\nonumber\\
 & & \Big[|0,0,0,A_+\rangle+|0,A_+,0,0\rangle\nonumber\\
 &  & +|A_+,A_q,A_{-q},0\rangle+|A_+,0,A_q,A_{-q}\rangle\Big]\nonumber\\
 &  & -\frac{1}{\sqrt{12}}
M^{(n)}(2,P,1)\times\nonumber\\
 & & \Big[|Q_{1-q},0,0,A_q\rangle+|Q_{1-q},A_q,0,0\rangle\Big]\nonumber\\
\end{eqnarray}
\begin{eqnarray}
\Sigma_+^{(n)}(P_1)|3,0\rangle & = &
-\frac{1}{\sqrt{6}}
M^{(n)}(1,P,0)\times\nonumber\\
 & & \Big[|0,0,A_+,0\rangle+|A_+,A_q,0,A_{-q}\rangle\Big]\nonumber\\
 & & -\frac{1}{\sqrt{6}}
M^{(n)}(2,P,1)|Q_{1-q},0,A_q,0\rangle\nonumber\\
 & & \\
\Sigma_+^{(n)}(P_1)|4,0\rangle & = &
-\frac{1}{3}
M^{(n)}(1,P,0)|0,A_q,A_+,A_{-q}\rangle\nonumber\\
 & & -\frac{1}{3}
M^{(n)}(2,P,1)|Q_{1-p},A_q,A_p,A_{-q}\rangle\nonumber\\
 & & \\
\Sigma_+^{(n)}(P_1)|5',0\rangle & = & -2
M^{(n)}(1,P,0)\times\nonumber\\
 &  & \Big[|0,A_q,A_{-q},A_+\rangle+|0,A_+,A_q,A_{-q}\rangle\Big]\nonumber\\
 & & -2
M^{(n)}(2,P,1)\times\nonumber\\
 & & \Big[|Q_{1+q},A_{-p},A_p,A_{-q}\rangle\nonumber\\
 & & +|Q_{1+q},A_{-q},A_{-p},A_{p}\rangle\Big]\nonumber\\
\end{eqnarray}
\begin{eqnarray}
\Sigma_+^{(n)}(P_1)|6,0\rangle & = & 
-\frac{1}{2\sqrt{10}}
M^{(n)}(1,P,0)\times\nonumber\\
 & & \Big[|0,0,Q_{1-q},A_q\rangle+|0,A_q,Q_{1-q},0\rangle\nonumber\\
 & & +|0,A_q,0,Q_{1-q}\rangle+|0,Q_{1-q},0,A_q\rangle\nonumber\\
 & & +|A_+,Q_{-r},A_q,A_{p}\rangle+|A_+,A_r,A_q,Q_{-r}\rangle\Big]\nonumber
\end{eqnarray}
\begin{eqnarray}
 & & -\frac{1}{2\sqrt{10}}
M^{(n)}(2,P,1)\times\nonumber\\
 & & \Big[-|A_{1-r},A_q,A_p,0\rangle_2-|A_{1-r},0,A_p,A_q\rangle_2\nonumber\\
 & & +|Q_{1+p},0,Q_{-r},A_{-q}\rangle+|Q_{1+q},A_p,0,Q_{-r}\rangle\nonumber\\
 & & +|Q_{1+p},A_q,Q_{-r},0\rangle+|Q_{1+q},Q_{-r},0,A_{p}\rangle\Big]\nonumber\\
\end{eqnarray}
\begin{eqnarray}
\Sigma_+^{(n)}(P_1)|7,0\rangle & = &
-\frac{1}{\sqrt{20}}
M^{(n)}(1,P,0)\times\nonumber\\
 & & \Big[|0,Q_{1-p},A_p,0\rangle+|0,0,A_q,Q_{1-q}\rangle\nonumber\\
 & & +|A_+,A_q,Q_{-r},A_p\rangle\Big]\nonumber\\
 & & -\frac{1}{\sqrt{20}}
M^{(n)}(2,P,1)\times\nonumber\\
 & & \Big[-|A_{1-p-q},A_p,0,A_{q}\rangle_2\nonumber\\
 & & +|Q_{1+q},Q_{-r},A_p,0\rangle\nonumber\\
 & & +|Q_{1+p},0,A_q,Q_{-r}\rangle\Big]\nonumber\\
\label{S+7_0}
\end{eqnarray}
It is convenient to express these states in a new basis which defines
the occupation of the plaquettes $P_1$ $P_2$ $P_3$ $P_4$ in cyclic order, e.g.
\begin{eqnarray}
\Sigma_+^{(n)}(P_1)|1,0\rangle & = & -M^{(n)}(1,P,0)|A_+,0,0,0\rangle
\end{eqnarray}
where
\begin{eqnarray}
-M^{(n)}(1,P,0) & = & \langle A_+,0,0,0|\Sigma_+(P_1)|1,0\rangle
\label{M1P0}\\
|A_+,0,0,0\rangle & = & \left(\begin{array}{cc}
0 & 0\\
A_+ & 0
\end{array}\right)\label{A000}
\end{eqnarray}
In this way, we are led to the following triplet states:
\begin{eqnarray}
|A_+,A_q,A_{-q},0\rangle & = & \sum_q(-)^q
\left(\begin{array}{cc}
A_q & A_{-q}\\
A_+ & 0\end{array}\right)\label{AAA0_1}
\end{eqnarray}
\begin{eqnarray}
|Q_{1-q},0,0,A_q\rangle & = & \sum_q(-)^q
\left(\begin{array}{cc}
0 & 0\\
Q_{1-q} & A_q\end{array}\right)\nonumber\\
 & & \hspace{0.4cm}\times\left(\begin{array}{c|cc}2 & 1 & 1\\
1-q & 1 & -q\end{array}\right)
\end{eqnarray}
\begin{eqnarray}
|Q_{1-p},A_q,A_p,A_{-q}\rangle & = & \sum_{p,q}
(-)^{p+q}\left(\begin{array}{cc}
A_q & A_p\\
Q_{1-p} & A_{-q}\end{array}\right)\nonumber\\
 & & \hspace{0.4cm}\times\left(\begin{array}{c|cc}2 & 1 & 1\\
1-p & 1 & -p\end{array}\right)
\end{eqnarray}
\begin{eqnarray}
|A_+,Q_{-r},A_q,A_p\rangle & = & \sum_{p,q}
(-)^{p+q}\left(\begin{array}{c|cc}2 & 1 & 1\\
r & q & p\end{array}\right)\nonumber\\
 & & \hspace{0.4cm}\times\left(\begin{array}{cc}
Q_{-r} & A_q\\
A_+ & A_p\end{array}\right)
\end{eqnarray}
\begin{eqnarray}
 & & |A_{1+p},0,Q_{-r},A_q\rangle=\sum_{p,q}
(-)^{p+q}\times\nonumber\\
 & & \left(\begin{array}{c|cc}2 & 1 & 1\\
r & p & q\end{array}\right)
\left(\begin{array}{cc}
0 & Q_{-r}\\
A_{1+p} & A_q\end{array}\right)
\left(\begin{array}{c|cc}2 & 1 & 1\\
1+p & 1 & p\end{array}\right)\nonumber\\
\end{eqnarray}
\begin{eqnarray}
 & & |A_{1-r},A_q,A_{p},0\rangle_2=\sum_{p,q}(-)^{p+q}\times\nonumber\\
 & & \left(\begin{array}{c|cc}2 & 1 & 1\\r & q & p\end{array}\right)
\left(\begin{array}{c|cc}2 & 1 & 1\\-r & -1 & 1-r\end{array}\right)
\left(\begin{array}{cc}A_q & A_p\\A_{1-r} & 0\end{array}\right)\nonumber\\
\label{AAA0_2}
\end{eqnarray}
\begin{eqnarray}
 & & |Q_{1+p},0,Q_{-r},A_q\rangle=\sum_{p,q}(-)^{p+q}\times\nonumber\\
 & & \left(\begin{array}{c|cc}2 & 1 & 1\\r & q & p\end{array}\right)
\left(\begin{array}{c|cc}2 & 1 & 1\\1+p & 1 & p\end{array}\right)
\left(\begin{array}{cc}0 & Q_{-r}\\Q_{p+1} & A_q\end{array}\right)\nonumber\\
\label{Q0QA}
\end{eqnarray}
The transition matrix elements induced by staggered spin operators
$\Sigma_+^{(n)}(P_1)$ are expressed by the reduced matrix elements
$M^{(n)}(1,P,0)$ (\ref{M1P0}) and $M^{(n)}(2,P,1)$ and appropriate
Clebsch-Gordan coefficients; the latter are absorbed in the definitions
(\ref{A000})-(\ref{Q0QA}) of the triplet states.
Apart from rotations, there are two different states (\ref{AAA0_1}),
(\ref{AAA0_2}) with three triplets. They turn out to be orthogonal
to each other:
\begin{eqnarray}
\langle A_+,A_q,A_{-q},0|A_{1-r},A_q,A_p,0\rangle & = & \nonumber\\
=\sum_q(-)^q
\left(\begin{array}{c|cc}2 & 1 & 1\\0 & q & -q\end{array}\right)
\left(\begin{array}{c|cc}2 & 1 & 1\\0 & -1 & 1\end{array}\right) & = & 0\nonumber\\
 &  & \label{sc_AAA0}
\end{eqnarray}

\item[b)]
In this step we rotate the triplet states (\ref{S+1_0})-(\ref{S+7_0}) and
(\ref{A000})-(\ref{Q0QA}) as described in (\ref{rotation}).
In this way we obtain the decompositions (\ref{decomposition_1}).

\item[d)]
In this step we compute the scalar products (\ref{decomposition_2}). The $n$-independent
$7\times 7$ matrix (\ref{rrec_1}) is given as

\begin{eqnarray}
\Gamma(\gamma) & = & \left(\begin{array}{cc}
\Gamma(5,5) & \Gamma(5,6) \\
\Gamma^T(5,6) & \Gamma(6,6)
\end{array}\right),\quad
i,i'=1,2,3,4,5\nonumber\\
 & & \label{Gamma1}
\end{eqnarray}
with
\begin{eqnarray}
\Gamma(5,5) & = & \frac{1}{4}\left(\begin{array}{ccccc}
1 & \frac{1}{\sqrt{3}} & \frac{1}{\sqrt{6}} & 0 & 0\\
\frac{1}{\sqrt{3}} & 1+\frac{\tilde Y}{6} & \frac{\sqrt{2}}{3} &
\frac{1}{3\sqrt{3}} & \frac{7+\frac{1}{3}}{\sqrt{60}}\\
\frac{1}{\sqrt{6}} & \frac{\sqrt{2}}{3} & \frac{4+\tilde Y}{6} &
\frac{1}{\sqrt{6}} & \frac{1}{\sqrt{30}}\\
0 & \frac{1}{3\sqrt{3}} & \frac{1}{\sqrt{6}} & \frac{1+\tilde Y}{3} &
2\frac{1+\tilde Y}{3\sqrt{20}}\\
0 & \frac{7+\frac{1}{3}}{\sqrt{60}} & \frac{1}{\sqrt{30}} &
2\frac{1+\tilde Y}{3\sqrt{20}} & \frac{7}{5}(1+\tilde Y)
\end{array}\right),\nonumber\\
 & & \hspace{1.0cm}\tilde Y=\gamma^2 Y\nonumber\\
 & & \\
%
\Gamma^T(5,6) & = & \frac{\gamma}{4}\left(\begin{array}{ccccc}
0 & \frac{Y}{2\sqrt{30}} & \frac{Y}{2\sqrt{15}} &
\frac{Y}{3\sqrt{10}} & \frac{1}{10\sqrt{2}}(1+\frac{4}{3}Y)\\
0 & \frac{Y}{2\sqrt{15}} & 0 &
\frac{Y}{6\sqrt{5}} & \frac{1}{6}Y\\
\end{array}\right),\nonumber\\
 & & \\
%
\Gamma(6,6) & = & \frac{1}{80}\Bigg[
\left(\begin{array}{cc}
x_1 & 0 \\
0 & x_2 
\end{array}\right)\nonumber\\
 & & \hspace{0.5cm}+\sqrt{2}\Big(2Y+\gamma^2(X+Z)\Big)
\left(\begin{array}{cc}
0 & 1 \\
1 & 0
\end{array}\right)\Bigg]\,.\nonumber\\
 & & \quad\label{Gamma4}
\end{eqnarray}
with:\,
\begin{eqnarray}
x_1 & = & 3Y+10+\gamma^2(2Y+3Z)\\
x_2 & = & 2Y+5+\gamma^2(Y+2Z)\\
 & & X=\frac{16}{9},\,\,\,Y=\frac{5}{3},\,\,\,Z=\frac{41}{18}.
\end{eqnarray}

\end{itemize}

\end{appendix}







\begin{thebibliography}{90}

\bibitem{bednorz86}J.G. Bednorz, K.A. M\"uller,
Z. Phys. B\,{\bf 64}, 189 (1986)

\bibitem{hubbard63}J. Hubbard, Proc. Roy. Soc. {\bf A}\,276, 238 (1963)

\bibitem{anderson87}P.W. Anderson, Science {\bf 235}, 1196 (1987)

\bibitem{baskaran87}G. Baskaran, Z. Zou, P.W. Anderson, 
Sol. State Comm. {\bf 63}, 973 (1987)


\bibitem{schulz92}H.J. Schulz, T.A.L. Ziman,
Europhys. Lett. {\bf 18}, 355 (1992)

\bibitem{manousakis91}E. Manousakis, Rev. Mod.
Phys. {\bf 63}, 1 (1991)

\bibitem{lin92}H.Q. Lin, D.K. Campbell, 
Phys. Rev. Lett. {\bf 69}, 2415 (1992)

\bibitem{lin94}H.Q. Lin, D.K. Campbell, Y.C. Cheng, C.Y. Pan,
Phys. Rev. B\,{\bf 50}, 12702 (1994)

\bibitem{chakravarty88}S. Chakravarty, B.I. Halperin, D.R. Nelson,
Phys. Rev. Lett. {\bf 60}, 1057 (1988); Phys. Rev. B\,{\bf 39} (1989)
%

\bibitem{fisher89} D.S. Fisher,
Phys. Rev. B\,{\bf 39}, 11783 (1989)

\bibitem{affleck86}I. Affleck,
Phys. Rev. Lett. {\bf 56}, 746 (1986)

\bibitem{affleck87}I. Affleck, F.D.M. Haldane,
Phys. Rev. B\,{\bf 36}, 5291 (1987)

\bibitem{hamer94} C. J. Hamer, W. Zheng, and J. Oitmaa,
Phys. Rev. B 50, 6877 (1994)


\bibitem{huse88}D.A. Huse,
Phys. Rev. B\,{\bf 37}, 2380 (1988)

\bibitem{sandvik97}A. W. Sandvik,
Phys. Rev. B\,{\bf 56}, 11678 (1997)

\bibitem{loew07}U. L\"ow,
Phys. Rev. B\,{\bf 76}, 220409 (2007)

\bibitem{edmonds60}A.R. Edmonds, ``Angular Momentum in Quantum Mechanics'',
Princeton University Press (1960)

\bibitem{tang89}S. Tang, J.E. Hirsch,
Phys. Rev. B\,{\bf 39}, 4548 (1989)

\bibitem{bernu92}B. Bernu, C. Lhuillier, L. Pierre,
Phys. Rev. Lett.\,{\bf 69}, 2590 (1992)

\bibitem{manousakis93}M.-B. Lepetit, E. Manousakis,
Phys. Rev. B\,{\bf 48}, 1028 (1993)

\bibitem{capponi04}S. Capponi, A. L\"auchli, M. Mambrini,
Phys. Rev. B {\bf 70}, 104424 (2004);
S. Capponi, Theor. Chem. Acc. {\bf 116}, 524 (2006)

\bibitem{troyer08}A. F. Albuquerque, M. Troyer, J. Oitmaa,
Phys. Rev. B\,{\bf 78}, 132402 (2008)

\bibitem{morningstar94}C. J. Morningstar and M. Weinstein,
Phys. Rev. Lett. {\bf 73}, 1873 (1994);
Phys. Rev. D {\bf 54}, 4131 (1996)

\bibitem{af08b}A. Fledderjohann, A. Kl\"umper, K.-H. M\"utter,
subm. to Eur. Phys. J. B, 2008



%
%



%

%




%














































































































\vspace{0.5cm}




\end{thebibliography}
\end{document}